\documentclass[twocolumn,
preprintnumbers,amsmath,amssymb,prb]{revtex4}
\usepackage[dvips]{color}

\usepackage{graphicx}
\usepackage{dcolumn}
\usepackage{bm}
\usepackage{latexsym,amsmath,amssymb,amsfonts,amsthm,enumerate}

\newcommand{\nn}{\nonumber}
\newcommand{\bs}{\boldsymbol}
\newcommand{\la}{\langle}
\newcommand{\ra}{\rangle}
\newcommand{\mef}{m_{\text{eff}}}

\newcommand{\eq}[1]{
\begin{equation}
#1
\end{equation}} 

\newcommand{\eqq}[2][]{
\begin{equation}
#2 \label{#1}
\end{equation}} 

\newcommand{\eem}[2][]{
\begin{align}
#2 \label{#1}
\end{align}}

\newcommand{\ee}[1]{
\begin{align}
#1
\end{align}}

\begin{document}


\title{Quantum quench in interacting field theory: \protect\\ A self-consistent approximation}

\author{Spyros~Sotiriadis}
\affiliation{SISSA and INFN, Sezione di Trieste, via Beirut 2/4, I-34151, Trieste, Italy}
\author{John~Cardy}
\affiliation{Rudolf Peierls Centre for Theoretical Physics, 1 Keble Road, Oxford OX1 3NP, UK \\
All Souls College, High Street, Oxford OX1 4AL, UK}

\date{\today} 

\begin{abstract}
We study a composite quantum quench of the energy gap and the interactions in the interacting $\phi^4$ model using a self-consistent approximation. Firstly we review results for free theories where a quantum quench of the energy gap or mass leads for long times to stationary behaviour with thermal characteristics. An exception to this rule is the $2d$ case with zero mass after the quench. In the composite quench however we find that the effect of the interactions in our approximation is simply to effectively change the value of the mass. This means on the one hand that the interacting model also exhibits the same stationary behaviour and on the other hand that this is now true even for the massless $2d$ case. 
\end{abstract}


\maketitle



\section{Introduction}
\label{secdef}

An area that has been gaining increasing interest over the last years is that of out-of-equilibrium quantum physics. An example of particular simplicity is that of quantum quenches in which some of the parameters of the hamiltonian of an isolated quantum system are changed instantaneously. Then one practically has to study the time evolution of a trial wavefunction, which is typically the ground state of the hamiltonian before the quench, under the influence of the hamiltonian after the quench. Although one expects a periodic collapse and revival of the initial state, in practice this period diverges rapidly with the system size and for large systems local observables may exhibit stationary behaviour at long times, eventhough the global wavefunction itself may never become such. This has been shown to be the case in many different settings \cite{rig-06,rig-07,caz-06,cc-06,cc-07,l-07,gdlp-07,kol-08,bs-08,qr-08,ke-08,fcmse-08,cfmse-08,bpgda-09,ro-09,fm-09}.

An obvious interesting question is whether this stationary behaviour is thermal as one may reasonably expect. It turns out that in many integrable systems the stationary behaviour is described by a statistical distribution which is similar to but not exactly thermal \cite{rig-06,rig-07,caz-06,cc-07,ke-08,fm-09}. 
More precisely it is a generalized Gibbs ensemble, subject to the constraints imposed by the integrals of motion. It was then conjectured that non-integrability is responsible for the exact thermalization of a system \cite{rig-08}. To what extend this is true is however still under investigation, since theoretical arguments that support this conjecture are based on semiclassical conjectures \cite{sr-94}, while numerical studies\cite{kol-07,mwnm-07,ro-09,rig-08,rig-09,bkl-09} lead to rather controversial results: some of them \cite{kol-07,mwnm-07,ro-09} reveal non-thermal behaviour even for non-integrable systems, while others \cite{rig-08,rig-09} are in good agreement with the thermal predictions and attribute the previous disagreement to finite size effects. There are some analytical studies in lattice models too \cite{qr-08,ek-08}: the first \cite{qr-08} refers to the Bose-Hubbard model but after the quench the system evolves under the free hamiltonian of the superfluid regime. In the second \cite{ek-08} dynamical mean-field theory (DMFT) applied to the non-integrable Falicov-Kimball model shows non-thermal features. 
On the other hand, an interaction quench in the Fermi-Hubbard model is possible to lead to thermalization, as shown using two different analytical \cite{mk-08} and numerical DMFT \cite{ekw-09} approximations.

In the present work we will study quantum quenches employing a field theoretic approach, which is supposed to capture their essential general characteristics. We consider systems described by a relativistic dispersion relation with some energy gap (or mass) and a maximum group velocity of excitations. Then for free systems, a quantum quench of the energy gap leads to stationary behaviour and a momentum dependent effective temperature can be defined \cite{cc-06,cc-07}. This is true for quite general conditions: the energy gap after the quench must be nonzero in $1d$ and $2d$ while in $3d$ the result holds even if it is zero. Furthermore a $1d$ gapless system can only be interacting and it turns out that it exhibits similar behaviour too \cite{cc-06,cc-07}. It should be emphasized that the notion of effective thermalization used throughout the present and some related earlier work\cite{cc-07} refers to the thermal-like stationary behaviour that fits to the generalized Gibbs ensemble description rather than the standard thermal theory. This is manifest in the fact that each momentum mode corresponds to a different effective temperature, since in the absence of interactions each mode evolves independently from the others. This suggests that in interacting systems the energy exchange due to collisions between different momentum modes would result in a mixing of their effective temperatures. However if the interaction is such that the system is still integrable then there will be some other decomposition into independent modes (quasiparticles) and we expect that the system still exhibits stationary behaviour with a different effective  temperature for each of these modes. Therefore it is only when the interaction makes the system non-integrable that thermalization to a unique common temperature is still a possibility.

The simplest interacting field theory is the $\phi^4$ model. We consider a simultaneous quench of the mass from $m_0$ to $m$ and of the coupling constant from $\lambda_0$ to $\lambda$. In order to study the evolution of the system we need to use an approximation scheme and the simplest one is the Hartree-Fock or self-consistent approximation. This can be applied in a number of different but equivalent ways. In perturbation theory it consists in ignoring all skeleton diagrams from which the diagrammatic expansions of correlation functions are constructed, except for the simplest one, i.e. the loop diagram. This turns out to be the same as approximating the system's state by gaussian wavefunctions or substituting the quartic interaction term in the hamiltonian by a quadratic one with a self-consistent coefficient. Notice however that in this simple approximation, collisions between particles of different momenta are neglected and this makes our approach incapable of answering the previous question about the relation between non-integrability and exact thermalization. Indeed, although the $\phi^4$ model is non-integrable, the Hartree-Fock approximation becomes exact only in the large-$N$ limit of the linear $\sigma$-model, i.e. the generalization of the $\phi^4$ model to an $N$-component field, which becomes integrable in this limit. Thus our approach provides the integrable counterpart of a non-integrable model that best approximates it. It is however the necessary first step towards understanding the effect of quantum quenches in interacting systems and should be expected to reveal some of their general features.

There is a significant number of publications that use the same method to study other closely related out-of-equilibrium problems, partially due to applications to cosmology. Cooper and Mottola \cite{como-87} make a detailed presentation of the method for the evolution of a general trial wavefunction and Boyanovsky et al. \cite{bo-93,bls-93,bdv-93} study the special case of a quench from the disordered to the ordered phase at large temperature and in $3d$. Also Wetterich et al. have studied the time evolution of out-of-equilibrium initial ensembles using a different method based on the numerical computation of the time-dependent effective action\cite{bw-98,bw-99}. Recently a remarkable numerical study based on the same method and including next-to-leading order effects in the large-$N$ expansion, has shown that an initial pure state evolves so that the reduced density matrix indeed thermalizes at large times\cite{gs-09}.

Using our approximation we find that the two point correlation function long after the quench is of the same form as the free correlation function but with a different mass that has to be determined self-consistently. This means that nothing really changes in terms of the relaxation of the system: once again it becomes stationary and a momentum dependent effective temperature can be defined, the only difference being that $m$ will be replaced by an effective mass $m^*$ which depends also on the coupling constant $\lambda$. The self-consistency equation for $m^*$ has always a real solution larger or equal to $m$. In the critical case $m=0$, we find that in $1d$ $m^*$ is also zero, but in $2d$ it becomes finite. This leads to the important conclusion that in $2d$, after a quench to zero mass which according to the above discussion would \emph{not} lead to relaxation if the system were free, now due to the presence of the interaction, it acquires a non-zero effective mass which allows it to relax. Furthermore, by studying the time evolution of the effective mass we find that if $m_0>m$ and $\lambda$ is sufficiently large then right after the quench the system is effectively set into an unstable state although it soon recovers its stability.

In the first part of this paper we focus on free systems which have been partially discussed earlier \cite{cc-06,cc-07}. Here we present an elegant simplified derivation of the quench propagator, develop an exact imaginary time formulation based on an earlier invented mapping to a slab geometry and define an average measure of the effective temperature first introduced in recent work \cite{scc-09}. These constitute a useful toolkit for many applications and extensions. For completeness we briefly report earlier results regarding $1d$ integrable systems with critical evolution. In the second part we study the composite quench in the $\phi^4$ model in the self-consistent approximation. This part is split into two sections: in the first we follow a heuristic approach based on perturbation theory and find an ansatz for the correlation function and in the second we start with the equations of motion and investigate the time evolution to verify the results obtained from our ansatz.

\section{Simple harmonic oscillator}

The simplest problem of a quantum quench one can start with is that of a simple harmonic oscillator whose frequency is quenched from $\omega_0$ to $\omega$. The hamiltonian before the quench is 
\eq{H_0=\frac{1}{2} \pi^2 +  \frac{1}{2} \omega^2_0 \phi^2 }
while after the quench it is 
\eq{H=\frac{1}{2} \pi^2 +  \frac{1}{2} \omega^2 \phi^2 }
The initial state is the ground state $|\Psi_0\ra$ of $H_0$.

From a physical point of view, what happens is that $|\Psi_0\ra$, as a trial state different from the ground state $|0\ra$ of $H$, contains, compared to that, an energy excess which is distributed to the excitation levels of $H$. After the quench the evolution of the wavefunction in the Schr\"{o}dinger picture is given by
\eqq[evol]{|\Psi(t)\ra = e^{-iHt}|\Psi_0\ra = \sum{e^{-i(n+1/2)\omega t} |n\ra\la n|\Psi_0\ra}}
where $|n\ra$ is an arbitrary eigenstate of $H$. 

It is trivial to observe that the evolution is periodic since after a period $T=2\pi/\omega$ the system returns back to the initial state, up to an irrelevant minus sign. This is a special case of quantum recurrence \cite{quant_rec}. In fact the wavefunction will exhibit periodicity or quasi-periodicity (i.e. it will return arbitrarily close to the initial state after sufficiently large time) in any system with \emph{discrete} energy eigenvalues. Systems with finite degrees of freedom always have such discrete spectra, while in the thermodynamic limit the spectrum becomes in general continuous and quantum recurrence may be lost. In practice even for finite but large systems, the corresponding period is usually so large that this periodicity is irrelevant.

\subsection{Propagator}
\label{secFFTprop}

We are also interested in the correlation function of the field operator $\phi$ at different times, i.e. the propagator $\langle \Psi_0 | \mathcal{T}\{\phi(t_1) \phi(t_2)\}| \Psi_0\rangle \equiv C_q(t_1,t_2)$ where $\mathcal{T}$ denotes time ordering. The time evolution of $x$ in the Heisenberg picture is given by the equations of motion 
\eq{\ddot{\phi}+\omega^2 \phi = 0} 
which can be solved easily
\eqq{\phi(t)=\phi(0)\cos \omega t + \pi(0)\frac{\sin \omega t}{\omega} }
We therefore have
\begin{multline}
\langle \Psi_0 | \phi(t_1) \phi(t_2) | \Psi_0\rangle = \la \Psi_0|\phi^2(0)|\Psi_0\ra \cos \omega t_1 \cos \omega t_2 + \\ 
+ \la \Psi_0|\pi^2(0)|\Psi_0\ra \frac{\sin \omega t_1 \sin \omega t_2}{\omega^2} + \\
+ \la \Psi_0| \phi(0) \pi(0) + \pi(0) \phi(0) |\Psi_0\ra \frac{\sin \omega (t_1+t_2)}{2 \omega} - \\ 
- i \frac{\sin \omega (t_1-t_2)}{2 \omega}\label{gen_prop}
\end{multline}
where the canonical commutation relation $[\phi(0),\pi(0)]=i$ has been taken into account in order to simplify the last term. All terms are symmetric under the interchange  $t_1 \leftrightarrow t_2$ apart from the last one which is antisymmetric. Thus time ordering amounts to substituting $(t_1-t_2)$ in the last term by its absolute value. 

It is now clear that the problem reduces to the calculation of the initial expectation values of $\phi^2(0)$, $\pi^2(0)$ and $\phi(0)\pi(0)+\pi(0)\phi(0)$. From the initial condition that the system lies in the ground state of $H_0$ we easily find 
\begin{subequations}
\label{inexpval}
\begin{align}
& \la \Psi_0| \phi^2 (0)|\Psi_0\ra = \frac{1}{2 \omega_0} \\
& \la \Psi_0| \pi^2 (0)|\Psi_0\ra = \frac{\omega_0}{2} \\
& \la \Psi_0| \phi(0) \pi(0) + \pi(0) \phi(0) |\Psi_0\ra =0 
\end{align}
\end{subequations}
and substituting into (\ref{gen_prop}) we obtain
\eem[q-prop]{C_q(t_1,t_2) = & \frac{(\omega -\omega_0)^2}{4 \omega^2 \omega_0} \cos \omega (t_1-t_2)+ \nn \\
& + \frac{\omega^2 -\omega_0^2}{4 \omega^2 \omega_0} \cos \omega (t_1+t_2)+ \frac{1}{2 \omega} e^{-i \omega|t_1-t_2|}}
Notice that we have separated the Feynman propagator $e^{-i \omega |t_1-t_2|}/2 \omega$ which, as expected, is the only term that survives if $\omega=\omega_0$ i.e. if there is no quench at all. Also notice that the only term that breaks time invariance is the second one.

\section{Linearly coupled oscillators (free fields)}

Let us now move on to study a system of linearly coupled harmonic oscillators or equivalently a free field theory. In general such a system is described by a quadratic hamiltonian of the form 
\eq{H=\frac{1}{2} \sum_r{ \pi^2(r) + \frac{1}{2} \sum_{r,r'} K(r-r')( \phi(r) - \phi{(r')})^2}}
which can be easily diagonalised in momentum space where it takes the form 
\eqq[Hk]{H=\sum_k{\frac{1}{2} \pi_k \pi_{-k} +  \frac{1}{2} \omega^2_k \phi_k \phi_{-k} }}
We will assume a relativistic dispersion relation
\eqq[disp1]{\omega^2_k = c^2 k^2 + m^2 c^4}
with energy gap (or mass, in the language of quantum field theory) $m$ and speed of sound $c$. This can also describe successfully non-relativistic systems with the same energy gap $m$ and maximum velocity of excitations $c$. 

The quantum quench that we will consider consists in an instantaneous change of the mass from $m_0$ to $m$. For brevity we can set $c=1$. An investigation of a quench of the speed of sound $c$ is done elsewhere\cite{scc-09}. As earlier, we assume that before the quench at $t=0$ the system lies in the ground state of the initial hamiltonian $|\Psi_0\ra$. In addition the system is kept isolated from the environment before and after the quench.

In order to study the time evolution, it is sufficient to find the two-point correlation function i.e. the propagator
\eq{ \la \Psi_0 |\mathcal T \{\phi(r_1,t_1)\phi(r_2,t_2)\}|\Psi_0\ra \equiv C_q(t_1,t_2,r_1-r_2) }
since in a free theory all physical observables can be obtained from this. From (\ref{Hk}) we see that the system is decomposed into a set of independent momentum modes each of which evolves as a simple harmonic oscillator. Thus the propagator is simply the Fourier transform with respect to $k$ of the expression (\ref{q-prop}) with $\omega_{0k} = \sqrt{k^2+m_0^2}$ and $\omega_k = \sqrt{k^2+m^2}$
\eqq[q-prop2]{C_q(t_1,t_2, r) = \int{\frac{d^d k}{(2 \pi)^d} \; e^{i \bs{k} \cdot \bs{r}} C_q(t_1,t_2;k)}}

\subsection{Properties of the propagator}
\label{secpropprop}

Let us now study the physical properties of the equal time propagator in real space. For simplicity we will mainly use its asymptotic form for $m_0 \gg m$ and $t,r \gg m_0^{-1}$. This will be called the \emph{deep quench limit} and should obviously exhibit all characteristic features of a quantum quench since it is one of the two most extreme possibilities for the relation between the two masses. In this limit the propagator simplifies to
\eqq[integral1]{C_{dq}(r,t) = \int{\frac{d^d k}{(2 \pi)^d} \; e^{i \bs{k \cdot r}} \frac{m_0}{4 \omega_k^2} (1-\cos 2 \omega_k t) }}
The massless ($m=0$) and massive ($m\neq 0$) cases are different and should be investigated separately. 

\subsubsection{Massless case}

In this case $\omega_k = |k|$ and after some algebra using Fourier transforms of common functions, we obtain the following exact results:
\begin{itemize}
	\item $d=1$
\eqq[1dprop]{C_{dq}^{(1d)}(r,t) = 
\begin{cases}
0 \quad &\text{ if } r>2t, \\
m_0 (2 t -r) /8 \quad &\text{ if } r<2t.
\end{cases}}
	\item $d=2$
\eq{C_{dq}^{(2d)}(r,t) = 
\begin{cases}
0 \quad &\text{ if } r>2t, \\
\frac{m_0 }{8 \pi} \log [({2 t+\sqrt{4 t^2-r^2}})/{r}] \quad &\text{ if } r<2t.
\end{cases}}
	\item $d=3$
\eq{C_{dq}^{(3d)}(r,t) =
\begin{cases}
0 \quad &\text{ if } r>2t, \\
m_0 /16 \pi r \quad &\text{ if } r<2t.
\end{cases}}
\end{itemize}
In all dimensions we distinguish between two spacetime regions in which the behaviour of the propagator is qualitatively different: for $r>2t$ it is always zero, unlike for $r<2t$. This means that the correlations between two points at distance $r$ remain unchanged until $t=r/2$. Also notice that in $3d$ the propagator is time independent for $r<2t$.

\subsubsection{Massive case}

By evaluating the integral (\ref{integral1}) we notice that, as before, we have to distinguish between two spacetime regions in which the behaviour of the propagator is qualitatively different. If $r>2t$ then we can close the integration contour in the upper half of the complex $k$-plane and since there is no pole the integral is zero. If on the other hand $r<2t$ then for the time independent part of the integrand we close the integration contour in the upper half plane but for the time dependent part we have to rotate it by $90^\circ$ instead. Each part has single poles at $k=\pm i m$ and the outcome is nontrivial. Exact results cannot be found and we have to employ asymptotic methods for large $r$ and $t$. In particular using the stationary phase method we find that for fixed $r$ and large $t$ the time dependent part of the integral tends to zero like $t^{-d/2} \cos{2 m t}$. Also the rest decreases for large $r$ like $e^{-mr}/r^{(d-1)/2}$. 

\begin{figure}[htbp]
	\centering
		\includegraphics[width=1.00\columnwidth]{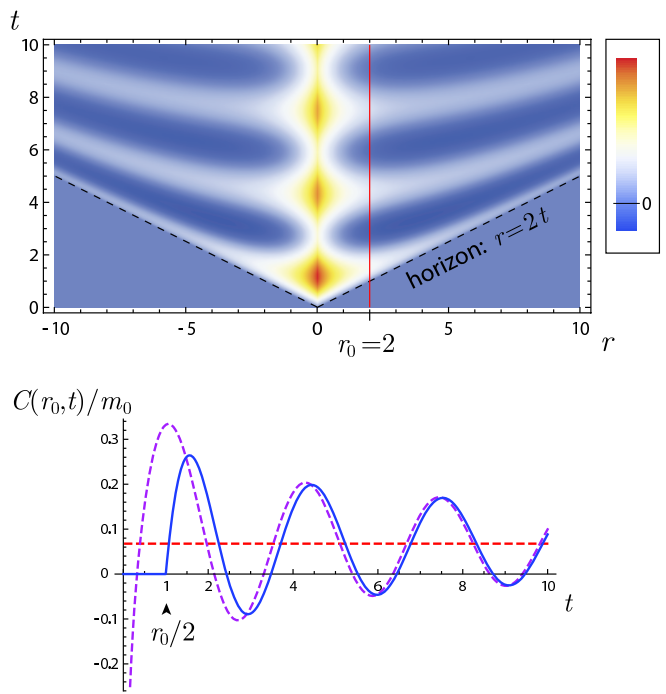}
	\caption{\emph{Top}: Spacetime plot of the deep quench propagator $C_{dq}(r,t)$ in $1d$ and for $m=1$, as obtained by numerical integration of (\ref{integral1}). The horizon effect is clearly demonstrated. Outside the horizon the value is exactly zero. \emph{Bottom}: Time dependence of $C_{dq}(r,t)$	(blue line) at fixed distance $r=r_0=2$, denoted by the vertical red line in the above figure. The dashed lines give the large time asymptotic expressions. Notice the decaying oscillations $\sim t^{-1/2} \cos{2 m t}$ (purple line) around the stationary value $\sim e^{-mr}$ (red line). }
	\label{fig:horizon}
\end{figure}

We thus conclude that the propagator changes sharply as we cross the lines $t=r/2$. Before this time there are no correlations between two distant points, while afterwards the two points become correlated. This feature, which is a direct consequence of the causality principle, is called the \emph{horizon effect}. Fig.~\ref{fig:horizon} illustrates the main features of the massive propagator in $1d$. 

Another particularly important conclusion is that if $m \neq 0$ then for fixed distance the propagator becomes \emph{stationary} for large times. The same is true for $m=0$ in $3d$, but not in $1d$ or $2d$. In addition this result is robust and does not rely on the deep quench approximation. Indeed if we use the full expression of $C_q(t;k)$ (\ref{q-prop}) for $m=0$ and $3d$ we find that the time dependence decays exponentially. The $1d$ case is more complex and requires special treatment. We will talk about this in section \ref{massless}.

\subsection{Comparison with the slab propagator}
\label{secslab}

\begin{figure}[htbp]
	\centering
	\includegraphics[width=.70\columnwidth]{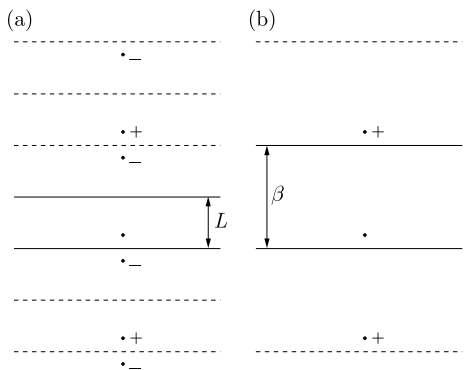}
	\caption{Images required for the slab with Dirichlet (a) or periodic (b) boundary conditions.}
	\label{fig:slab}
\end{figure}

We will now study a completely different problem which however turns out to be an imaginary time formulation of a quantum quench. We consider a euclidean free field theory defined on a $(d+1)$-dimensional slab of thickness $L$ with Dirichlet boundary conditions, that is the two-point correlation function or Green's function vanishes when one of the points are on the boundaries of the slab $\tau=-L/2$ and $\tau=+L/2$, where $\tau$ is the transverse coordinate. 
The Green's function $G_{sl}({r_1},\tau_1,{r_2},\tau_2)$ for this problem can be found using the method of images as follows: to reproduce the boundary conditions we put an infinite set of alternating positive and negative `charges' at the reflections of the `source' on the boundaries  (Fig.~\ref{fig:slab}.a). 
Then $G_{sl}({r}_1,\tau_1,{r}_2,\tau_2)$ is the superposition of (euclidean) Feynman propagators between $({r}_2,\tau_2)$ and each of the images of the source $({r}_1,\tau_1)$. Since the problem is translationally invariant in the $d$ longitudinal directions, in the mixed $({k},\tau)$ representation we find that $G_{sl}(\tau_1,\tau_2;k)$ is 
\ee{
& \frac{1}{2 \omega_k} \left( {\sum\limits_{n=0}^\infty{e^{-\omega_k(|\tau_1-\tau_2|+2nL)}} + \sum\limits_{n=1}^\infty{e^{-\omega_k(-|\tau_1-\tau_2|+2nL)}} }\right.- \nn \\ 
& \left.{- \sum\limits_{n=0}^\infty{e^{-\omega_k\left (\tau_1+\tau_2+(2n+1)L\right )}} - \sum\limits_{n=1}^\infty{e^{-\omega_k\left (-\tau_1-\tau_2+(2n-1)L\right )}}} \right) 
}
This is a geometric series and the result is
\eem[slab-prop]{
& \frac{{e^{-\omega_k |\tau_1-\tau_2|} + e^{+\omega_k (|\tau_1-\tau_2|-2L)} - 2 e^{-\omega_k L}\cosh \omega_k (\tau_1+\tau_2)}}{2 \omega_k (1-e^{-2 \omega_k L})} \nn \\
& = \frac{\cosh \omega_k (\tau_1-\tau_2) }{\omega_k (e^{2\omega_k L}-1)}  - \frac{e^{\omega_k L} \cosh \omega_k (\tau_1+\tau_2)}{\omega_k (e^{2\omega_k L}-1)} + \nn \\
& + \frac{1}{2\omega_k} e^{-\omega_k|\tau_1-\tau_2|}
}
By analytically continuing to real times $\tau \to i t$ we find
\begin{multline}
\label{slab-prop2} 
 G_{sl}(t_1,t_2;k) = \\
 = \frac{\cos \omega_k (t_1-t_2) }{\omega_k (e^{2\omega_k L}-1)} - \frac{e^{\omega_k L} \cos \omega_k (t_1+t_2)}{\omega_k (e^{2\omega_k L}-1)}  + \frac{1}{2\omega_k} e^{-i\omega_k|t_1-t_2|} 
\end{multline}
If we now compare the slab propagator in real time (\ref{slab-prop2}) with the quench propagator (\ref{q-prop}) we notice that these are exactly equal if and only if
\begin{subequations}
\label{conditions}
\begin{align}
 \frac{(\omega_{0k}-\omega_k)^2}{4 \omega_k \omega_{0k}} & = \frac{ 1}{e^{2 \omega_k L}-1} \\
 \frac{\omega_{0k}^2-\omega_k^2}{4 \omega_k \omega_{0k}} & = \frac{ e^{\omega_k L}}{e^{2 \omega_k L}-1}
\end{align}
\end{subequations}
Remarkably, the above two conditions are consistent and the solution is
\ee{\tanh{(\omega_k L/2)} = 
\begin{cases}
{\omega_k}/{\omega_{0k}} \qquad \text{ if } \omega_k<\omega_{0k}, \\
{\omega_{0k}}/{\omega_{k}} \qquad \text{ if } \omega_k>\omega_{0k}.
\end{cases}}
Notice that if we solve with respect to $L$, the answer is a function of $k$. 

Thus the problem of a quantum quench can be equivalently formulated as a euclidean theory on a slab with momentum dependent thickness. The initial conditions in real time are translated into boundary conditions on the slab. In the deep quench limit $m_0\to\infty$ the condition becomes $L \sim 2/m_0 $, independent of $k$ and therefore the analogy between the quantum quench and the slab is asymptotically exact. The reason is that Dirichlet boundary conditions correspond to vanishing initial value of the quench propagator, which is indeed the case for $m_0\to \infty$, since $C_q(0,0;k)=1/2 \omega_{0k} \to 0$. 

It should be mentioned that our choice of Dirichlet boundary conditions has nothing special: in fact it is only important in the deep quench limit. One can verify that the quench propagator can be similarly identified with the Green's function corresponding to the following general boundary conditions (known as Robin or `impedance' boundary conditions due to their application to electromagnetics)
\eq{a {\hat{G}_{sl}(\tau_1,\tau_2;k)} + b \frac{\partial {\hat{G}_{sl}(\tau_1,\tau_2;k)}}{\partial n} = 0 }
where 
\ee{& a = \omega_k \sinh(\omega_k/\omega_{0k}) - \omega_{0k} \cosh(\omega_k/\omega_{0k}) \nn \\
& b = \cosh(\omega_k/\omega_{0k}) - \omega_{0k}/\omega_k \sinh(\omega_k/\omega_{0k})}
$\partial /\partial n$ denotes the normal derivative at the boundary $\tau = \pm L/2$ and in this case $L$ is chosen to be $L=2/\omega_{0k}$. Note that for $\omega_{0k} \gg \omega_k$ the latter condition reduces to Dirichlet type. To intuitively understand the meaning of these boundary conditions, we can use an analogy from electromagnetics. There, Dirichlet boundary conditions correspond to complete reflection by a perfect conductor, while Robin boundary conditions correspond to partial reflection and refraction by an imperfect conductor with a large refractive index. 

The correspondence between a quantum quench and the slab construction turns out to be valid, at least in the deep quench limit, even in interacting theories where an exact solution may not be possible\cite{cc-07}.

\subsection{Comparison with the thermal propagator}
\label{sec:compth}

Let us now compare the above two propagators with the thermal or \emph{Matsubara} propagator, which describes a system at thermal equilibrium at finite (inverse) temperature $\beta$. As is well-known, in imaginary time this corresponds to the Green's function in the geometry of a $(d+1)$-dimensional cylinder of circumference $\beta$, i.e. a slab of equal thickness with periodic instead of Dirichlet boundary conditions. Among other ways, this can also be derived using the method of images. To reproduce the periodic boundary conditions we now need to put only the positive images (Fig.~\ref{fig:slab}.b) and the result is 
\eqq[ther-prop]{ G_{th}(\tau_1,\tau_2;k)= \frac{1}{2 \omega_k} \left({e^{-\omega_k |\tau_1-\tau_2|} + 2 \frac{\cosh \omega_k (\tau_1-\tau_2)}{e^{\beta \omega_k}-1}}\right)}
or in real time, after the analytical continuation $\tau\to i t$ 
\eqq[ther-prop2]{ G_{th}(t_1,t_2;k)= \frac{1}{2 \omega_k} \left({e^{-i\omega_k |t_1-t_2|} + 2 \frac{\cos \omega_k (t_1-t_2)}{e^{\beta \omega_k}-1}}\right)}

We observe that if we could ignore the $(t_1+t_2)$-dependent part then the slab propagator $G_{sl}(t_1,t_2;k)$ and the quench propagator $C_q(t_1,t_2;k)$  would be the same as the thermal propagator $G_{th}(t_1,t_2;k)$ with $L=\beta/2$. This can actually be correct for the real space form of the quench propagator at large times, as we have already seen in section \ref{secpropprop}. Indeed this is the case if $m\neq 0$ or $m=0$ and $d=3$. 

As a conclusion, at large times the system tends to a state with thermal-like correlation functions, which is what we named \emph{effective thermalization}. The effective temperature is given, according to all the above, by the condition
\ee{\tanh{(\beta_{\text{eff}} \omega_k /4)} = 
\begin{cases}
{\omega_k}/{\omega_{0k}} \qquad \text{ if } \omega_k<\omega_{0k}, \\
{\omega_{0k}}/{\omega_{k}} \qquad \text{ if } \omega_k>\omega_{0k}.
\end{cases}}
Notice that the effective temperature is momentum dependent, which could be expected since, as we already mentioned, in a free system each momentum mode evolves independently from the others and there is no reason why they should all thermalize to the same temperature. Yet in the deep quench limit the effective temperature becomes momentum independent $\beta_{\text{eff}}\sim 4/m_0$. 

It should be emphasized that the state itself is neither thermal nor stationary: the density operator still exhibits oscillating behaviour for example. However, since in a free system all local observables can be derived from the two-point correlation function which does become stationary, the same happens to all such observables as well. It is crucial that the system is in the \emph{thermodynamic limit} and the observables under consideration are \emph{local} since then an integration over an infinite set of momenta is required and it is exactly this interference of all independent momentum modes that leads to thermalization. Such observables include those defined on any finite subsystem $A$ of the whole system, like the reduced density operator of $A$ \cite{p-03}. In this sense the complement of $A$ acts as a thermal bath with which $A$ comes into thermal equilibrium. This explains why the effective thermalization that we consider does not contradict with the fact that in a free or more generally integrable system, there is an infinite set of conserved quantities that prevent the system from thermalizing as a whole. The subsystem $A$ is not closed and there are no such restrictions to prevent its thermalization.

\subsection{Estimation of the effective temperature from the field fluctuations}
\label{avetemp}

As we saw, the effective temperature in our free model is different for each momentum mode. Since the low momentum modes are those that determine the large distance behaviour, for most purposes $\beta_{\text{eff}}(k=0)$ is sufficient in order to macroscopically describe the system. We can define  \cite{scc-09} 
however an estimate of the effective temperature that averages over all momentum modes in a natural way, by comparing the field fluctuations long after the quench $\langle \phi^2(x=0,{t\to\infty}) \rangle$ with those of a system at thermal equilibrium. We can call this \emph{average effective temperature} and denote it as $\bar{\beta}$. Then $\bar{\beta}$ must satisfy
\eqq[avtemp]{\int{d^dk \; {C}^*_{q}(k;m,m_0)} = \int{d^dk \; G_{th}(k;m,\bar\beta)}}
where ${C}^*_q$ stands for the stationary part of the quench propagator. More explicitly
\eq{\int_0^\infty{k^{d-1} dk \frac{(\omega_{0k}-\omega_k)^2}{4 \omega_{0k} \omega_k^2}} = 
\int_0^\infty{k^{d-1} dk \frac{1}{\omega_k (e^{\bar\beta \omega_k}-1)}} }
from which we can find $\bar{\beta}$ as a function of $m$ and $m_0$. 
The latter can be written in dimensionless form as
\eq{m_0^{d-1} f_d(m/m_0) = m^{d-1} g_d(\bar\beta m) }
where 
\eqq[int1]{f_d(s) = \int\limits_0^\infty{k^{d-1} dk \; \frac{(\sqrt{k^2+1}-\sqrt{k^2+s^2})^2}{4 \sqrt{k^2+1} ({k^2+s^2})}}}
and 
\eqq[int2]{g_d(s) = \int\limits_0^\infty{k^{d-1} dk \; \frac{1}{\sqrt{k^2+1} (e^{s \sqrt{k^2+1}} - 1)}}}
In units of $m_0$, setting $x=m/m_0$ and $y=\bar\beta m_0$ we have
\eq{{x^{d-1} = \frac{f_d(x)}{g_d(xy)}}}
Tables \ref{tab1} and \ref{tab2} show the asymptotic behaviour of the integrals in several limits for the relation between the parameters and when possible their exact form. 
\begin{table*}[htbp]
\caption{Asymptotic behaviour of the quench integral $f_d(s)$~(\ref{int1})}
\label{tab1}
\begin{center}
\begin{tabular}{|c|l|l|l|l|}
\hline
$d$ & exact & $s\approx 0$ & $s\approx 1$ & $s\to \infty$ \\
\hline
1 & $ [ 2 \log s +({\sqrt{1-s^2}}/{s}) \, \arccos s ]/4 $ & $ ({\pi}/{2 s} + 2 \log s )/4$ & $(s-1)^2/6$ & $(\log s)/4$ \\
2 & $ [ 2 (s-1)- \sqrt{s^2-1} \, \arccos (1/s) ]/4 $ & $ -(\log s)/4 $ & $(s-1)^2/12$ & $(1+{\pi}/{4})s/2$ \\
3 & $ [ {(1-s^2)}/{2}-s^2 \log s- s \sqrt{1-s^2} \, \arccos s ]/4 $ & $(1-\pi s)/8$ & $(s-1)^2/12$ & $(\log 2 - {1}/{2}) s^2/4$ \\
\hline
\end{tabular}
\end{center}
\end{table*}

\begin{figure}[htbp]
\begin{center}
\includegraphics[width=1.00\columnwidth]{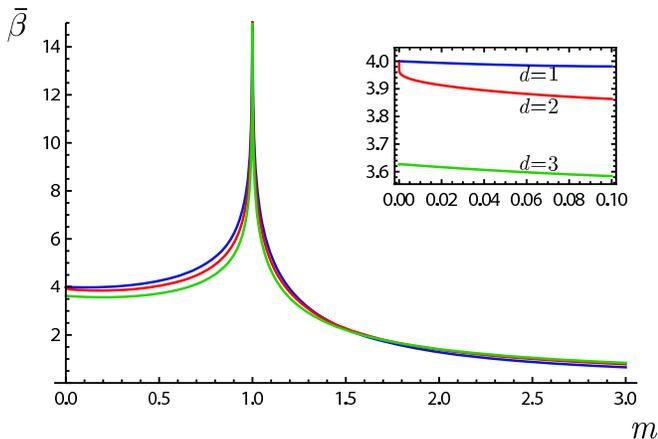}
\caption{Effective temperature as a function of the final mass $\bar\beta m_0 = F_d (m/m_0)$ in units of the initial mass $m_0=1$. Inset: Asymptotic behaviour for small $m$. Notice the logarithmic corrections in $2d$.}
\label{fig:1}
\end{center}
\end{figure}

\begin{table*}[htbp]
\caption{Asymptotic behaviour of the thermal integral $g_d(s)$~(\ref{int2})}
\label{tab2}
\begin{center}
\begin{tabular}{|c|c|l|l|}
\hline
$d$ & exact & $s\approx 0$ & $s\to \infty$\\
\hline
1 & -- & $ ({\pi}/{2 s}) + (\log s)/2 $ & $e^{-s}\sqrt{\pi/2s} $ \\
2 & $-\log(1-e^{-s}){/s}$ & $-{(\log s)/s} $ & $ {e^{-s}/s}$ \\
3 & -- & ${({\pi^2}/{6 s^2})[1-({3s}/{\pi})]}$ & ${e^{-s}\sqrt{\pi/2} s^{-3/2}}$\\
\hline
\end{tabular}
\end{center}
\end{table*}
Figure \ref{fig:1} shows a plot of $\bar\beta$ as a function of $m$ in units of $m_0$ as obtained numerically from the above equation. Note that for $m = m_0$ i.e. no quench at all, the effective temperature $1/\bar\beta$ is zero as it should be. Apparently in the deep quench limit the small wavelength behaviour dominates and according to an earlier comment in section \ref{sec:compth} 
we expect to find $\bar\beta \sim m_0^{-1}$ for any dimension. For small values of $m/m_0$ the asymptotic expressions of $f_d$ and $g_d$ allow us to calculate analytically the first order corrections of $\bar\beta$ as a function of $m/m_0$. In this way we find 
\begin{itemize}
	\item $d=1$
\eq{\bar\beta = \frac{4}{m_0}+\frac{32 \log 2 m}{\pi m_0^2} + ...} 
	\item $d=2$
\eq{\bar\beta = \frac{4}{m_0}\left (1 + \frac{3 \log 2 -2}{\log(m/m_0)} + ... \right )} 
	\item $d=3$
\ee{\bar\beta & = \frac{1}{m_0}\left({2 \pi}/{\sqrt{3}} - \pi (2-\pi/\sqrt{3}){m}/{m_0}+...\right) \nn \\
& \approx m_0^{-1}(3.6276-0.584967 m/m_0 + ...)}
\end{itemize}
In $1d$ and $2d$ the $k=0$ momentum mode dominates so that the first order term is $\bar\beta = 4/m_0$. In $2d$ however the logarithmic corrections could render comparison with data difficult. In $3d$ the contribution of nonzero but small $k$ modes causes a small shift of the numerical factor from 4 to ${2 \pi}/{\sqrt{3}} \approx 3.6276$.

\section{Massless $1d$ theories}
\label{massless}

In section \ref{secpropprop} we saw that for $d=1$ and $m=0$ the propagator does not become stationary. However the situation is different when we consider a physical 1$d$ quantum system, for the following reason. A massless free theory is not physically meaningful. The infrared divergences impose the introduction of interaction counterterms of all orders in perturbation theory over the introduced coupling constant. The field renormalization finally results in the physical field defined as the exponential of the original gaussian field $\phi$ (the vertex operator). Thus in a physically meaningful 1$d$ system, interaction terms must always be present and the correlation function is given by the expectation values of vertex operators $\la e^{i q\phi(x)} e^{-i q\phi(x')}\ra$ for an appropriate value of the constant $q$. This can be evaluated readily using the well-known property of gaussian integrals 
\eq{ \langle e^{iq\phi(x)} e^{-iq\phi(x')} \rangle = e^{-q^2\langle(\phi(x)-\phi(x'))^{2}\rangle/2} = e^{-q^2(C(0)-C(x-x'))} } 
where $C(x-x')=\la\phi(x)\phi(x')\ra$ is the free propagator we have already found. From (\ref{1dprop}) we obtain
\eqq[cft1]{\langle e^{iq\phi(0,t)} e^{-iq\phi(r,t)} \rangle = 
\begin{cases}
e^{-q^2 m_0 t/4} \qquad \text{ if } r>2 t, \\
e^{-q^2 m_0 r/8} \qquad \text{ if } r<2 t.
\end{cases}} 
Thus the linearly increasing time dependence of $C(r,t)$ leads to an exponentially decaying correlation function outside of the horizon and a static form inside the horizon. Therefore thermalization also occurs in 1$d$ systems. This has been shown to be the case for any massive to massless quench on a 1$d$ bosonic system, using the mapping to the slab and the powerful methods of conformal field theory \cite{cc-05,cc-06,cc-07}.

\section{Anharmonic coupled oscillators (interacting field theory) in self-consistent approximation.} \label{anh}

Let us now consider a system of anharmonic coupled oscillators. The simplest form of a hamiltonian describing such a system is 
\eqq[Hint]{H=\sum_r{\tfrac{1}{2} \pi^2 + \tfrac{1}{2} (\nabla \phi)^2 + \tfrac{1}{2} m^2 \phi^2 + \tfrac{1}{4!} \lambda \phi^4}}
In the continuum limit this corresponds to the simplest form of an interacting quantum field theory, the $\phi^4$ model. At $t=0$ we instantaneously change the mass from $m_0$ to $m$ and at the same time the coupling constant from $\lambda_0$ to $\lambda$. As before, we assume that initially the system lies in the ground state of the hamiltonian before the quench. 

Such a model is non-integrable and can be solved only approximately. In this paper we will focus solely on the Hartree-Fock or self-consistent approximation. Roughly speaking in this approach we assume that the quartic interactions can be approximated by a `mean field' quadratic term with a parameter that should be calculated self-consistently. More specifically the $\phi^4$ interaction term of the hamiltonian can be substituted as follows {\cite{ch-75}}
\eqq[subHF]{\phi^4 \to {-} 3 \la \phi^2\ra^2 + 6 \la \phi^2\ra \phi^2 }
where we have taken into account that $\la \phi \ra = 0$ and the numerical factors are derived by Wick's theorem as the number of combinations of operator contractions. 

Such a substitution is justified in the large-$N$ limit of the linear $\sigma$-model, which is a variant of the $\phi^4$ model where the field $\phi$ has $N$-components
\eqq[Hsigma]{H=\sum_{i=1}^N{\sum_r{\tfrac{1}{2} \pi_i^2 + \tfrac{1}{2} (\nabla \phi_i)^2 + \tfrac{1}{2} m^2 \phi_i^2 + \tfrac{1}{4!} \lambda (\phi^2_i)^2}}}
In the limit $N\to\infty$ with $\lambda N$ kept fixed, the Hartree-Fock approximation becomes exact.

Just by staring at (\ref{subHF}) we notice that the second term corresponds to a mass term but with a `mass' that has to be determined from the two point correlation function. The first term is just a number and does not affect the equations of motion but, as shown in appendix \ref{app:cons}, ensures the conservation of the total energy. Therefore we can define an effective mass $\mef$ according to
\eqq[self1]{\mef^{2} = m^2 + \frac{\lambda}{2} \sum_k{\la\phi_k^2\ra}}
and since the right hand side also depends on $\mef$, this is in fact a self-consistency equation for $\mef$. Note that the effective mass should correspond to the pole of the correlation function on the imaginary axis in the complex $k$-plane, which is what is physically measurable as the mass of the particles of the system. As a result of this approximation, our initial non-integrable problem has been effectively reduced to an integrable and in fact free one, subject to the self-consistency equation. In our out-of-equilibrium case we should keep in mind that the effective mass will be time dependent. 

We will use two slightly different methods in applying this approach. The first one is a perturbative method. After introducing the Schwinger-Keldysh method which is suitable for out of equilibrium problems, we soon realize that the usual perturbative expansion does not converge and a resummation of Feynman diagrams using the Dyson equation is needed. This leads us to a simple ansatz for the asymptotic form of the two point correlation function at large times. The second method emphasizes on the time evolution of the system and is based on a direct integration of the equations of motion in their, simplified by the self-consistent approximation, version. In order to solve these equations we employ an approximate analytical and an exact numerical method. The results of both calculations are in agreement with each other and additionally they verify our earlier ansatz. 

Before we start, it is worth to remind ourselves of the large-$N$ results for the ground state of our system, since in order to proceed to the out-of-equilibrium problem we will need to know more about the initial properties of the system. This will also introduce us to a discussion of the renormalization procedure and its application to the present problem. 

\subsection{Divergences and renormalization}\label{sec:div}

The initial two-point correlation function of our system in the large-$N$ limit is simply that of a free system with a mass equal to its effective value
\eqq[self1b]{{{\mef^{2}}_0} = m_0^2 + \frac{\lambda_0}{2} \sum_k{\la\phi_k^2\ra}}

The sum in the right hand side of (\ref{self1b}) represents the fluctuations of the field. In the continuum limit this corresponds to the integral
\eqq[ffluct]{\int{\frac{d^d k}{(2\pi)^d} \; \frac{1}{2 \sqrt{k^2+{\mef^2}_0}} }}
which exhibits ultraviolet (UV) divergences in all dimensions. In $1d$ and $2d$ these can be absorbed completely by a mass renormalization, while in $3d$ an additional coupling constant renormalization is required \cite{cjp74}. 

The mass renormalization amounts to allowing the bare mass $m_0$ to be divergent so as to compensate the divergent integral. The (finite) renormalized mass is defined by $m_{0R}^2 = m_0^2+\delta m_0^2$ where the mass counterterm $\delta m_0^2$ is 
\eqq[dm]{\delta m_0^2 = \frac{\lambda_0}{2} \int{\frac{d^d k}{(2\pi)^d} \; \frac{1}{2 \sqrt{k^2+m_{0R}^2}} }}
The effective mass in terms of $m_{0R}$ is then
\eem[self1c]{ & {\mef^{2}}_0 = m_{0R}^2 + \nn \\
& + \frac{\lambda_0}{2} \int{\frac{d^d k}{(2 \pi)^d} \left ( \frac{1}{2 \sqrt{k^2 + {\mef^2}_0}} - \frac{1}{2 \sqrt{k^2 + m_{0R}^2}} \right ) } }
which is finite in $1d$ and $2d$. 

In $3d$ there is still a logarithmic UV divergence in (\ref{self1c}) which can be absorbed by a coupling constant renormalization.  A suitable renormalization counterterm can be determined by studying the 4-point correlation function and turns out to be of the form 
\eqq[dl]{\delta \lambda_0 = \int{\frac{d^3 k}{(2 \pi)^3} \; \frac{1}{8 (k^2+m_{0R}^2)^{3/2}}}}
The resulting renormalized coupling constant $\lambda_{0R}$ satisfies 
\eqq[rencc]{\lambda_0 = \frac{\lambda_{0R}}{1-\lambda_{0R} \, \delta \lambda_0}}
and replacing in (\ref{self1c}) we obtain 
\eem[self1d]{{\mef^2}_0 & = m_{0R}^2 + \frac{\lambda_{0R}}{2} \int{\frac{d^3 k}{(2 \pi)^3} \; \left ( \frac{1}{2 \sqrt{k^2+{\mef^2}_0}}  - \right.}\nn \\ & {\left.-\frac{1}{2 \sqrt{k^2+m_{0R}^2}} + \frac{{\mef^2}_0 - m_{0R}^2}{4 (k^2+m_{0R}^2)^{3/2}}  \right )}}
which is indeed finite. Note that in all dimensions the solution to the above equations is  \eq{{\mef}_0 = m_{0R}} 
i.e. the renormalized mass is identical to the effective mass. In what follows we should keep in mind the well-known renormalization group result that in $3d$ the critical point of this model corresponds to zero coupling constant i.e. in the continuum limit the macroscopic behaviour of the theory is effectively free.
Therefore interactions are not physically meaningful in the continuum limit. In physical systems however, the existence of a finite lattice spacing that induces a natural UV cutoff renders all momentum integrals finite and there is not such a restriction.

After the quench, the change in the mass and the coupling constant results in a change of the corresponding counterterms as well. This is required, otherwise new divergences in the equation for the effective mass are inevitably born. On the other hand, this should not be regarded as a failure of renormalization theory as the latter does not have to apply to expectation values taken in states which are not obtained by renormalised operators acting on the vacuum and our initial state is not such. In absence of a definite rule for the selection of the renormalization counterterms like the one for the ground state or thermal expectation values, several choices can be applied\cite{como-87}. 
In the present work we will be using the ground state counterterms of the theory after the quench.

Since the field fluctuations right after the quench are exactly the same as before it, the equation for the effective mass right after the quench is 
\eqq[inmef0]{\mef^2(t\to 0^+) = m^2 +\frac{\lambda}{2} \int{\frac{d^d k}{(2 \pi)^d} \frac{1}{2 \sqrt{k^2 + m_{0R}^2}} } }
or introducing the mass renormalization
\eem[inmef1]{ & \mef^2(0^+) = m_R^2 + \nn \\
& + \frac{\lambda}{2} \int{\frac{d^d k}{(2 \pi)^d} \left ( \frac{1}{2 \sqrt{k^2 + m_{0R}^2}} - \frac{1}{2 \sqrt{k^2 + m_{R}^2}} \right ) } }
where $m_R$ is the renormalized mass after the quench and
\eq{\delta m^2 = \frac{\lambda}{2} \int{\frac{d^d k}{(2 \pi)^d} \frac{1}{2 \sqrt{k^2 + m_R^2}}}}
is the corresponding mass counterterm. As before the last expression is convergent in $1d$ and $2d$, but not in $3d$. If we use a coupling constant renormalization counterterm
\eqq[dl2]{\delta \lambda = \int{\frac{d^3 k}{(2 \pi)^3} \; \frac{1}{8 (k^2+m_{R}^2)^{3/2}}}}
we find 
\eem[mef3d]{\mef^2(& 0^+) = m_{R}^2 + \frac{\lambda_{R}}{2} \int{\frac{d^3 k}{(2 \pi)^3} \; \left ( \frac{1}{2 \sqrt{k^2+m_{0R}^2}}  - \right.}\nn \\ 
& {\left.-\frac{1}{2 \sqrt{k^2+m_{R}^2}} + \frac{\mef^2(0^+) - m_{R}^2}{4 (k^2+m_{R}^2)^{3/2}}  \right )}}
which is only convergent in the trivial case ${\mef(0^+) = m_{0R}}$ where there is no jump in the effective mass i.e. no quench at all. 
Recalling our previous remark, we realise that this problem is due to the fact that the presence of interactions does not make sense in the continuum limit. In lattice systems however there is not such a problem and the practical meaning of the above is simply that $\mef(0^+)$ is very large. In the following we will therefore keep a large UV cutoff $\Lambda$ in all expressions for the $3d$ case and investigate the dependence of our results on this.

An interesting first observation is that as defined by (\ref{inmef1}) the initial mass-square $\mef^2(0^+)$ can be negative. Indeed for $m_0<m$, $\mef^2(0^+)$ is always positive, but for $m_0>m$ the mass shift induced by the interactions is negative and if $\lambda$ is large enough then $\mef^2(0^+)<0$. In particular if $m=0$ the latter is always true. From a physical point of view this negativity means that the quench can effectively drag the system into an unstable initial state like that of a double well (or generally `mexican hat') potential. We will come back to this aspect of the problem later.

The integration in (\ref{inmef1}) can be done analytically. Expressing the integral in dimensionless form we have
\eq{\mef^2(0^+) = m_{R}^2 + \frac{\lambda_{R}}{2} \frac{\Omega_d}{(2\pi)^d} m_0^{d-1} h_d(m_R/m_0) }
where $\Omega_d$ is the total solid angle in $d$ dimensions ($\Omega_1 = 2, \Omega_2 = 2\pi, \Omega_3 = 4\pi$) and the function $h_d(s)$ is defined as
\eqq[hdef]{h_d(s) = \int\limits_0^{\Lambda\to\infty}{k^{d-1} dk \; \left ( \frac{1}{2 \sqrt{k^2+1}} - \frac{1}{2 \sqrt{k^2+s^2}}\right )}}
and can be easily shown to be
\eqq[hdef2]{ h_d(s) = 
\begin{cases}
(\log s)/2 \quad &\text{ if } d=1, \\
(s-1)/2 \quad &\text{ if } d=2, \\
(s^2-1) (\log \Lambda )/4 \quad &\text{ if } d=3.
\end{cases} }

\subsection{Perturbative approach}

\begin{figure}[htbp]
	\centering
	\includegraphics[width=1.00\columnwidth]{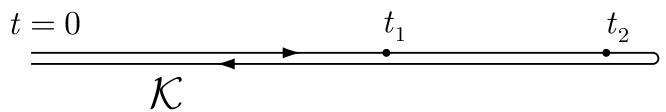}
	\caption{The Schwinger-Keldysh contour for a quantum quench.}
	\label{fig:Keldysh}
\end{figure}

In order to study the evolution of the effective mass for $t>0$ we have to calculate the two-point correlation function and the usual way to do this is to use perturbation theory. The two-point correlation function is
\eq{ \tilde{C}(r,t_1,t_2) \equiv \left\langle {\Psi _0 } \right|\phi (0,t_1 )\phi (r,t_2 )\left| {\Psi _0 } \right\rangle }
where, as before, $\left| {\Psi _0 } \right\rangle$ is the ground state of the initial hamiltonian. For simplicity let us first assume that $\lambda_0 =0$, i.e. that there is no interaction before the quench so that $\left| {\Psi _0 } \right\rangle$ is the ground state of a free hamiltonian. 
At this point we encounter an important difference with the usual QFT methods: 
at zero temperature the starting point of such a calculation is usually the following formula 
\eqq[qftcf]{
\frac{{\left\langle 0 \right|\mathcal{T}\{ \phi_i (0,t_1 )\phi_i (r,t_2 )\exp{\textstyle\left ( - i\int_{ - \infty }^{ + \infty } {dt\, H_{int}(t)} \right )}\} \left| 0 \right\rangle }}{{\left\langle 0 \right|\mathcal{T}\{ \exp{\textstyle\left ( - i\int_{ - \infty }^{ + \infty } {dt\, H_{int}(t)} \right )}\} \left| 0 \right\rangle }}}
But in our case this expression is inappropriate since it relies on the condition that $\left| 0 \right\rangle $ is the ground state of the free part of the hamiltonian and the interactions are swithed on and off adiabatically. In a quantum quench this is not valid because $\left| {\Psi _0 } \right\rangle $ is the ground state of a different hamiltonian and the changes are done instantaneously. Thus we have to trace back to the origin of (\ref{qftcf}) which follows from the interaction picture formalism
\eqq[Keldysh]{\left\langle {\Psi _0 } \right|\mathcal{T}\{ {\phi_i (0,t_1 ) \phi_i (r,t_2 )\exp{\textstyle\left ( - i \int_{\mathcal{K}} {dt \,H_{int}(t)}\right )}} \}\left| {\Psi _0 } \right\rangle  }
where $t$ is integrated over a contour $\mathcal{K}$ that starts from some initial time $t_i$, passes through $t_1$ and $t_2$ where the interaction picture field operators $\phi_i$ are placed, extends to some final time $t_f$ and then goes back to $t_i$ so that times on the second half of the contour are considered to be later than those on the first half (Fig.~\ref{fig:Keldysh}). 
This is the well-known Schwinger-Keldysh method for non-equilibrium quantum systems \cite{schw-61,keld-64,keld-65,chsu-85,calhu-88,kam-04,kam-09} and is applicable to any choice of initial state.

If $\left| {\Psi _0 } \right\rangle = \left| 0 \right\rangle $ and the interaction is switched on and off adiabatically then we can extend $t_i \to -\infty$ and $t_f \to +\infty$. In this case, from the adiabatic theorem, the action of $\exp{\textstyle ( + i\int_{ - \infty }^{ + \infty } {dt\, H_{int}(t)} )}$ (i.e. the evolution operator along the second half of the contour) on $\left| {\Psi _0 } \right\rangle$ yields just a multiplicative constant and (\ref{Keldysh}) reduces to (\ref{qftcf}). In the present problem we need to use the original expression (\ref{Keldysh}) instead. The initial time can be set to be $t_i=0$, when the interaction is switched on. However the same choice can be used even in the general case when $\lambda_0 \neq 0$, i.e. when the interaction is present before the quench, since as explained above, in our approximation the initial state is still that of a free theory but with the mass replaced by its effective value. 

It is worth to remark that an alternative way of deriving the Keldysh contour is by using the slab construction mentioned earlier. In this approach one would have to integrate in imaginary time from $-L/2$ to $+L/2$, i.e. from one to the other boundary of the slab, then analytically continue the arguments of the operators from imaginary to real times as in $\tau \to i t$ and finally take the limit $L \to 0$ thus recovering (\ref{Keldysh}). 

We can now expand (\ref{Keldysh}) in powers of $\lambda$. According to the above, the zeroth order perturbative term $\left\langle {\Psi _0 } \right|\phi_i (0,t_1 )\phi_i (r,t_2 )\left| {\Psi _0 } \right\rangle $ is exactly the quench propagator (\ref{q-prop2}) with the masses $m_0$ and $m$ replaced by their renormalized values. 
 
\begin{figure}[htbp]
	\centering
		\includegraphics[width=0.50\columnwidth]{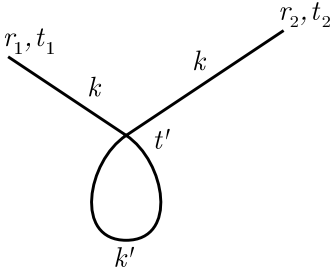}
	\caption{First order Feynman diagram.}
	\label{fig:loop}
\end{figure}

The first order correction $C^{(1)}(t_1,t_2;k)$ corresponds to the single loop Feynman diagram (Fig.~\ref{fig:loop}). 
After applying Wick's theorem we find that $C^{(1)}(t_1,t_2;k)$ reads 
\eqq[cf0]{{\frac{\lambda}{2}}\int_{\mathcal{K}}{dt' \; C(t_1,t';k)C(t_2,t';k)\int{\frac{d^d k'}{(2\pi)^d} \,C(t',t';k') }}}
The explicit form of the loop momentum integral $\int{d^d k \; C(t,t;k)}$ is
\eqq[lmi]{ \int{d^d k \; \left(\frac{{(\omega_{0k}  - \omega_k)^2 }}{{4\omega_k^2 \omega_{0k} }} - \frac{{m_0 ^2  - m^2}}{{4\omega_k^2 \omega_{0k} }}\cos 2\omega_k t + \frac{1}{2 \omega_{k}} \right)}}
but we also have to take into account the mass renormalization, which amounts to subtracting the UV divergent Feynman part and substituting the bare mass $m$ by the renormalized $m_R$
\eq{\int{d^d k \; C(t,t;k,m)} = \int{d^d k \; \left (C(t,t;k,m_R) - \frac{1}{2 \omega_{k,m_R}}\right ) }}
For brevity we redefine $m,m_0$ to be the renormalized masses $m_R,m_{0R}$ in all subsequent equations. Then (\ref{lmi}) can be written explicitly as
\eqq[expr1]{\int{d^d k \; \left(\frac{{(\omega_{0k}  - \omega_{k})^2 }}{{4\omega_{k}^2 \omega_{0k} }} - \frac{{m_0 ^2  - m^2}}{{4\omega_{k}^2 \omega_{0k} }}\cos 2\omega_{k} t\right)}}
From the terms that remain in (\ref{expr1}), the first one which is the time independent part is always convergent since it decays like $k^{-5}$ for large $k$. On the other hand, the second term which is the time dependent part decays like $k^{-3} \cos 2\omega_k t$, which means that it converges in $1d$ and $2d$, while in $3d$ it is divergent only at $t=0$. 
We also note that (\ref{expr1}) does not suffer from infrared (IR) divergences in the massless case $m=0$ except in $1d$. 

Having analyzed the convergence of the loop integral, let us now calculate it. If we assume that $m\neq 0$, then the time-independent part has been calculated exactly already in section \ref{avetemp}: it is equal (up to a numerical factor involving the total solid angle in $d$ dimensions) to $m_0^{d-1} f_d(m/m_0)$ where $f_d(s)$ is given in Table~\ref{tab1}. On the other hand we recall that the time-dependent part has been shown to decrease with time. More specifically using the stationary phase method we find that for large times it decays like
\eq{\frac{(m^2-m_0^2)m^{d-2}}{m_0}\frac{\cos(2 m t + \varphi)}{(m t)^{d/2}}}
However for small times, this same time-dependent part can be important (or even divergent as we saw that happens in $3d$). 

Thus we are naturally led to the question whether it is safe or not to completely ignore the time-dependent part of the loop integral in calculating $C^{(1)}$ for large times. If this is correct, then the effect of the loop diagram for large times is simply a mass shift equal to the time-independent part (recall that a mass renormalization counterterm induces a similar shift in the mass, but an infinite one). Higher orders in perturbation theory correspond to more loops and therefore one needs to employ a resummation of all orders in order to compute the actual mass shift. Such a resummation can lead to a non-perturbative dependence of the mass shift and the correlation function on the coupling constant. Indeed if we calculate $C^{(1)}$ from (\ref{cf0}) assuming that the loop integral is constant, then we find that the first order correction increases linearly with time, i.e. it will eventually become larger than the zeroth order term and therefore the perturbative series does not converge. The required resummation can be done using the Dyson equation as described in the next section.

\subsubsection{Resummation using the Dyson equation}

As well-known the Dyson equation is an integral equation satisfied by the two-point correlation function of an interacting theory that expresses the fact that the latter can be constructed from the propagator in a recursive fashion, using a number of `skeleton' diagrams as building blocks. In our problem and in the mixed representation the Dyson equation can be written in the form 
\eem[Dyson0]{& \tilde{C}(t_1,t_2;k)= \nn \\
& C(t_1,t_2;k)+\int_{\mathcal{K}}{dt' \int_{\mathcal{K}}{dt'' \; C(t_1,t';k) \Sigma(t',t'';k) \tilde{C}(t'',t_2;k)}}} 
where we denote the full correlation function by $\tilde{C}$ and $\Sigma$ is the self-energy insertion, i.e. a two-leg insertion also constructed recursively by the skeleton diagrams. 

\begin{figure}[htbp]
	\centering
		\includegraphics[width=1.0\columnwidth]{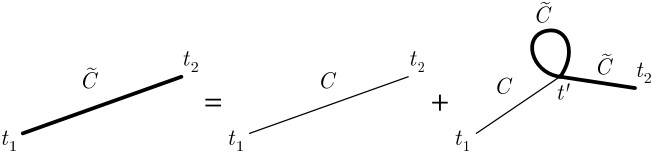}
	\caption{Diagrammatic representation of the Dyson equation in the large-$N$ limit.}
	\label{fig:Dyson}
\end{figure}

In the large-$N$ limit the loop diagram is the only skeleton diagram and therefore the self-energy is simply a loop of the full correlation function. The Dyson equation then takes the simplified form (Fig.~\ref{fig:Dyson})
\eem[Dyson]{& \tilde{C}(t_1,t_2;k)= \nn \\
& C(t_1,t_2;k)+\int_{\mathcal{K}}{dt'\; C(t_1,t';k) \Sigma(t') \tilde{C}(t',t_2;k)}}
where
\eqq[Dyson2]{\Sigma(t') = \frac{\lambda}{2} \int{\frac{d^d k'}{(2\pi)^d} \left (\tilde{C}(t',t';k') - \frac{1}{2 \omega_k}\right )}}
taking into account the mass renormalization. Notice that comparing with (\ref{self1}) we realize that $\Sigma(t)$ is nothing but the shift in the mass-square
\eqq[Dyson3]{\Sigma(t) = \mef^2(t) - m^2}

As we see the Dyson equation contains $\tilde{C}$ explicitly in the right hand side but also implicitly in the definition of $\Sigma$. Thus it is difficult to solve in general. In many cases it is useful as a check of validity for an ansatz: we assume a particular form for $\tilde{C}$, substitute in the Dyson equation and check the consistency or determine any free parameters. This is how we are going to use it in our problem. 

Let us therefore construct an ansatz based on the hypothesis that the time-dependence of the loop be negligible. Then the same can be assumed for the self-energy since this is nothing but a dressed loop, i.e. the sum of all `cactus-diagrams'. This would mean that $\Sigma(t)$ can be replaced by its large time stationary value $\Sigma^* = \lim_{t\to +\infty} \Sigma(t)$ or, according to (\ref{Dyson3}), that the effective mass itself can be considered as time-independent and equal to its large time stationary value $m^*= \lim_{t\to +\infty} \mef(t)$. In other words we suppose that the effective mass simply jumps at the time of the quench from $m_{0}$ to $m^*$ in which case the correlation function should simply be equal to the quench propagator for a quench from $m_{0}$ to $m^*$. Note that our assumption is twofold: first we assume that $\mef$ tends to a stationary value and second that this happens fast enough to approximate its evolution by a jump. 

According to the above, our ansatz is that the two point correlation function $\tilde{C}(t_1,t_2;k)$ is approximately the same as the propagator itself but with $m$ replaced by an asymptotic effective mass $m^*$
\eqq[ans]{\tilde{C}(t_1,t_2;k;m_0,m) \sim C(t_1,t_2;k;m_0,m^*)}
We expect this relation to be asymptotically exact for large times, when any memory of the initial evolution of the effective mass will have been lost.

Now that we have an ansatz for the correlation function we can use the Dyson equation to check its validity and determine the value of the free parameter $m^*$. By substituting (\ref{ans}) into (\ref{Dyson}) and (\ref{Dyson2}) and replacing $\Sigma(t)$ by $\Sigma^*$, we find that the Dyson equation is satisfied exactly when $m^{*}$ satisfies the self-consistency equation 
\eqq[self-cons0]{m^{*2}-m^2 = \Sigma^* = \frac{\lambda}{2} \int{\frac{d^d k}{(2\pi)^d} \left ({C}^*(k;m_0,m^*) - \frac{1}{2 \omega_k}\right )} }
where ${C}^*(k;m_0,m)$ is the stationary part of the propagator. 

One can also check whether the remainder of the large time asymptotic form of the Dyson equation 
\eq{\int_{\mathcal{K}}{dt'\, C(t_1,t';k;m_0,m) (\Sigma(t')-\Sigma^*) C(t',t_2;k;m_0,m^*)} }
with
\eq{\Sigma(t') = \frac{\lambda}{2} \int{\frac{d^d k}{(2\pi)^d} \left (C(t',t';k;m_0,m^*) - \frac{1}{2 \omega_k}\right )}}
tends to zero as supposed to. This is however a cumbersome calculation and will not be presented. We will later show an alternative way to study the time evolution and verify our ansatz, but for the moment let us focus on the self-consistency equation (\ref{self-cons0}) and investigate its solutions.

\subsubsection{Self-consistent calculation of the mass shift}

Written explicitly the self-consistency equation (\ref{self-cons0}) is 
\eqq[self-con]{{m^*}^2 = m^2 +  \frac{\lambda}{2} \int{\frac{d^d k}{(2 \pi)^d} \; \left (\frac{(\omega_{0k} - \omega^*_{k})^2}{4 \omega_{0k} {\omega_k^*}^2} + \frac{\omega_{k} - \omega^*_{k}}{2 \omega_k \omega_{k}^{*}} \right )}}
where $\omega_k^* = \sqrt{k^2 + {m^*}^2}$. 

Once again some comments about the 3$d$ case are due as (\ref{self-con}) contains a logarithmically divergent integral. Therefore a UV cutoff $\Lambda$ is assumed and the solutions $m^*$ will depend upon it. As can be verified however, the small $\lambda$ behaviour of $m^*$ is not affected by $\Lambda$. By the way the $\lambda$-counterterm (\ref{dl2}) would successfully remove the current divergence yielding the finite equation
\eem[self-con-3d]{{m^*}^2 = m^2 + & \frac{\lambda_R}{2} \int{\frac{d^3 k}{(2 \pi)^3} \; \left( \frac{(\omega_{0k} - \omega^*_{k})^2}{4 \omega_{0k} {\omega_k^*}^2} + \right. } \nn \\
& \left. + \frac{\omega_{k} - \omega^*_{k}}{2 \omega_k \omega_{k}^{*}}  + \frac{{m^*}^2 - m^2}{4 \omega_k^3}  \right) }
but according to the discussion in section \ref{sec:div}, this is supposed to be correct only for $\lambda \to 0$ and therefore provides no more information than (\ref{self-con}) with a cutoff. 

Going back to the general case, if we make the momentum integrals dimensionless then the self-consistency equation can be written as 
\eq{{m^*}^2 = m^2 + \frac{\lambda}{2} \frac{\Omega_d}{(2\pi)^d} \left [ m_0^{d-1}  f_d\left (\frac{m^*}{m_0}\right ) + {m^*}^{d-1} h_d\left (\frac{m}{m^*}\right ) \right ]}
where $f_d(s)$ and $h_d(s)$ are the previously defined functions (\ref{int1}) and (\ref{hdef}).

\begin{figure*}[htbp]
\begin{center}
\includegraphics[width=1.8\columnwidth]{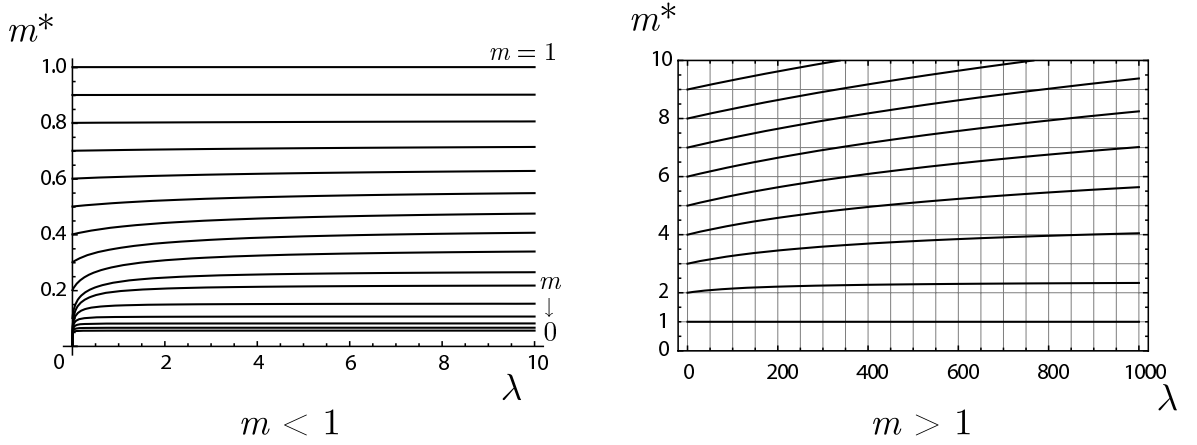}
\caption{Solutions of the self-consistency equation (\ref{self-con}) in $1d$. The plots show the effective mass $m^*$ as a function of the coupling constant $\lambda$ for several values of $m$ in units of $m_0=1$.  Notice that as $m\to 0$ the effective mass tends logarithmically to zero for all $\lambda$.}
\label{fig:2}
\end{center}
\end{figure*}

\begin{figure*}[htbp]
\begin{center}
\includegraphics[width=1.8\columnwidth]{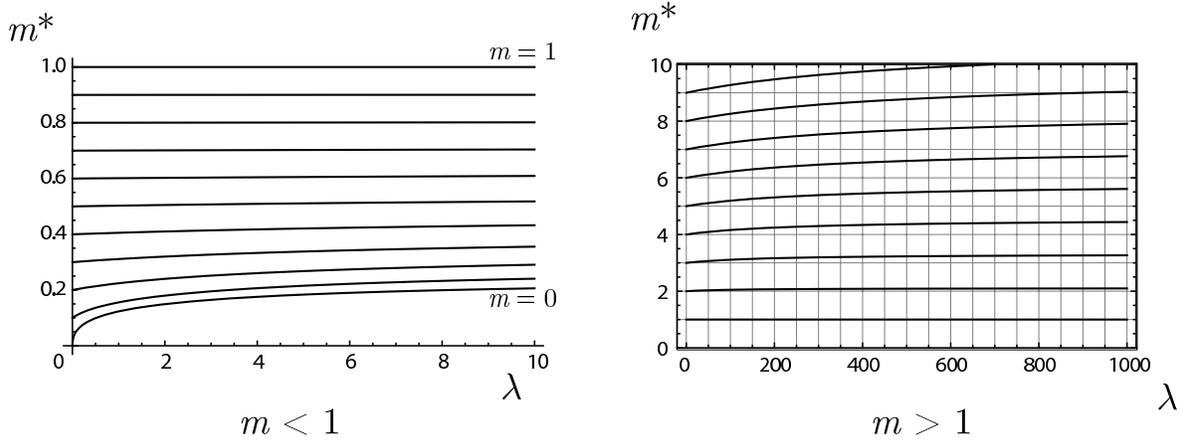}
\caption{The same plot in $2d$. Notice that, in contrast to the $1d$ case, as $m\to 0$ the effective mass tends to a non-zero value for all $\lambda>0$.}
\label{fig:3}
\end{center}
\end{figure*}

\begin{figure}[htbp]
\begin{center}
\includegraphics[width=.8\columnwidth]{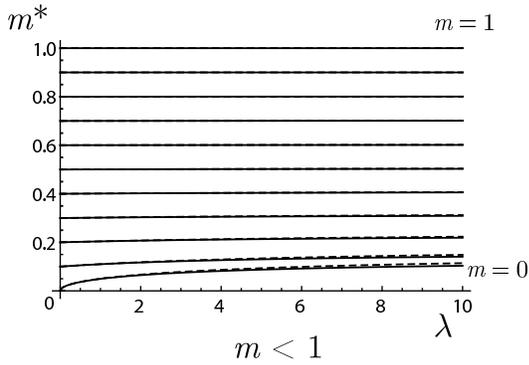}
\caption{The same plot in $3d$. The curves show a weak (but increasing for increasing $\lambda$) dependence on the cutoff $\Lambda$. The dashed lines correspond to $\Lambda=10^4$ while the full ones to $\Lambda=10^7$.}
\label{fig:4}
\end{center}
\end{figure}

\begin{figure}[htbp]
\begin{center}
\includegraphics[width=.9\columnwidth]{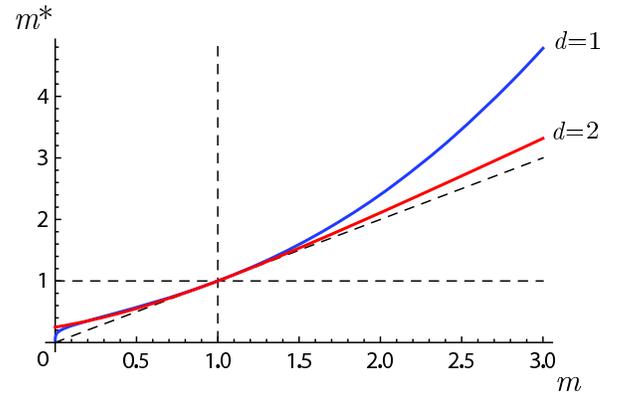}
\caption{Effective mass as a function of $m$ for $\lambda \to \infty$ in $1d$ (blue line) and $2d$ (red line) in units of $m_0=1$. The dashed straight lines are for reference. Notice that as $m\to 0$,  $m^*\to 0$ logarithmically in $1d$, while in $2d$ $m^* \to 0.24954$. }
\label{fig:3bs}
\end{center}
\end{figure}

The above equations can be solved numerically or even analytically in several asymptotic limits like for $\lambda\to 0$ or $m \to 0$. 
Fig.~\ref{fig:2},\ref{fig:3} and \ref{fig:4} show plots of the solutions  $m^*$ as a function of $\lambda$ for several values of $m$ in $1d$, $2d$ and $3d$, while Fig.~\ref{fig:3bs} shows $m^*$ as a function of $m$ for $\lambda\to\infty$ in $1d$ and $2d$. A first important remark is that for $m\neq 0$ and small $\lambda$ the first order correction $m^*-m$ is linear in $\lambda$, while for $m = 0$ this is not true. Instead $m^*$ depends on $\lambda$ in a non-perturbative way in this case. The first order corrections in $\lambda$ for $m=0$ are summarized below:
\begin{itemize}
	\item for $d=1$
\eq{m^* = 0 \quad \text{ for all } \quad \lambda}
In fact it is more correct to talk about the limit $m \to 0$, since $m$ can never reach zero in $1d$. In this limit, $m^*$ follows $m$ to zero like
\eq{m^* \sim \frac{m_0\pi/2}{2 \log (m_0/m) + 1 - 16 \pi^2 m^2/\lambda m_0^2}}
	\item for $d=2$
\eq{m^* = \frac{1}{4}\sqrt{ \frac{\lambda m_0}{2\pi} \log(m_0/\lambda)}}
	\item for $d=3$
\eq{m^* = \frac{m_0}{4 \pi \sqrt{2}}  \lambda^{1/2} }  
independent of the cutoff.
\end{itemize}

On the other hand for large $\lambda$ and $m$, $m^*$ increases like $m^*\sim m^2/2 m_0$ in $1d$, $m^*\sim 4 m/\pi$ in $2d$ while in $3d$ the large $\lambda$ result is cutoff dependent. In addition, in $2d$ and for $m=0$ and $\lambda \to \infty$ we find $m^* \to 0.24954... \, m_0$.

Of particular interest are the $2d$ results for $m=0$. The fact that $m^* \neq 0$ means that from the critical evolution in the presence of interactions, there always emerges a finite effective mass which lets the system become stationary, in contrast to the free case.

\subsection{Time evolution}

We saw in section \ref{sec:div} that the initial value of the effective mass-square $\mef^2(0^+)$ can be negative, while our ansatz suggests that its asymptotic final value is always positive. It is therefore worthwhile to investigate the time evolution of the effective mass in more detail. Although this can be done in the context of perturbation theory as in the previous section, an alternative and rather simpler way is by integrating the equations of motion for the field operator $\phi$. Since the exact equations are nonlinear, even if we were able to solve them the solution would depend on the initial operators $\phi(0),\dot \phi(0)$ in a nonlinear way, thus preventing a direct application of the initial conditions (\ref{inexpval}) as done in section \ref{secFFTprop}. Fortunately in the Hartree-Fock approximation this obstacle can be circumvented since the $\phi^4$ interaction term of the hamiltonian is substituted by a quadratic `mean field' term according to (\ref{subHF}). As explained in section \ref{anh}, this substitution reduces the interacting into a free problem with a time-dependent effective mass given by
\eqq[self1a]{\mef^{2}(t) = m^2 + \frac{\lambda}{2} \sum_k{\left (\la\phi_k^2(t)\ra-\frac{1}{2 \omega_k}\right )}}
thus yielding a linear equation of motion. 

Even after this simplification however the problem is not trivial. In the following two sections we will first apply an approximate method that leads to an analytical solution for small values of the coupling constant and later derive exact equations for the evolution of the correlation function which we will integrate numerically.

\subsubsection{Quasi-adiabatic self-consistent approximation}

A common approximation that could provide a completely analytical treatment is the adiabatic approximation which is based on the assumption that $\mef(t)$ varies slowly in comparison with the fast oscillations that characterize the solution \cite{calhu-88}. This is not a reasonable assumption though, since it is the solution itself that determines the time dependence of $\mef(t)$. However as we show below, one can establish an alternative argument leading to the same approximate solution. The latter becomes equivalent to our earlier ansatz (\ref{ans}) for small $\lambda$ and provides a first idea of the qualitative behaviour of the solution.

Since our problem is now free, it can once again be decomposed into a set of independent harmonic oscillators. Of course the time-dependence of the frequency of each oscillator involves a summation over the whole set of them, but for the moment it suffices to consider a single quantum harmonic oscillator with an arbitrary time-dependent frequency $\omega(t)$. The hamiltonian is
\eqq[timedepH]{H=\frac{1}{2}\pi^2 + \frac{1}{2}\omega^2(t) \phi^2}
The equation of motion for the field operator evolving under $\omega(t)$ is
\eqq[eqm]{\ddot{\phi} + \omega^2(t) \phi = 0}
If the frequency varies with time very slowly (adiabatically) then $\dot{\omega}/\omega^{2} \ll 1$ and as well-known the solution is given by
\eem[eq2]{\phi(t)= & \phi(0) \sqrt{\frac{\omega(0)}{{\omega(t)}}} \cos{\left ( \int_0^t{\omega(t') dt'}\right )} + \nn \\
& + \pi(0) \frac{1}{\sqrt{ \omega(t) \omega(0)}}  \sin{\left ( \int_0^t{\omega(t') dt'}\right )} }
A detailed derivation of the above equation in the quantum case can be found in appendix \ref{app:adiab}. 

Although, as we said, the adiabaticity condition does not apply to our problem because $\omega(t)$ may exhibit oscillations with the same frequency as the solution, the condition $\dot{\omega}/\omega^{2} \ll 1$ is also valid when the amplitude of the frequency oscillations is sufficiently small in comparison with the average value. This happens when the coupling constant $\lambda$ is sufficiently small so that from (\ref{self1a}) $\mef(t) \approx m$. 
In this quasi-adiabatic approximation we can still use the last expression (\ref{eq2}) as the solution to our problem. 

Having found the time evolution of $\phi$ we can use the initial conditions to derive the correlation function $\la \phi^2(t) \ra$ which is all we need in order to find $\mef(t)$. Recall that from (\ref{inexpval}) we have $ \la \phi(0)\pi(0) + \pi(0)\phi(0) \ra = 0$ and $ \la \phi^2(0) \ra = 1/2\omega_0$, $ \la \pi^2(0) \ra = \omega_0/2$. By a direct calculation
\eqq[a1]{\la \phi^2(t) \ra = {\frac{\omega^2(0) + \omega^2_0}{{4 \omega_0 \omega(t) \omega(0)}}} + \frac{\omega^2(0) - \omega^2_0}{{4 \omega_0 \omega(t) \omega(0)}} \cos\textstyle{\left (2\int_0^t{\omega_k(t') dt'}\right )}  }

Now going back to the interacting field theory model, we coclude that the equal time correlation function for each momentum mode $\la\phi_k^2(t)\ra$ is given by (\ref{a1}) with $\omega_k(t)$ corresponding to the time-dependent effective mass (\ref{self1a}) i.e. $\omega_k^2(t)=k^2+\mef^{2}(t)$. Therefore the self-consistency equation for $\mef(t)$ is
\eem[mt]{\mef^2 & (t) = m^2 + \frac{\lambda}{2} \sum_k \left[ {\frac{\omega_k^2(0) + \omega^2_{0k}}{{4 \omega_{0k} \omega_k(t) \omega_k(0)}} + } \right. \nn \\ 
& \left. + \frac{\omega_k^2(0) - \omega^2_{0k}}{{4 \omega_{0k} \omega_k(t) \omega_k(0)}} \cos\textstyle{\left (2\int_0^t{\omega_k(t') dt'}\right )} \displaystyle - \frac{1}{2 \omega_{k}}\right]}

This equation enables us to extract physical information about the evolution of the system through its only parameter $\mef(t)$. A first observation is that $\mef(t)$ depends on an average value over all previous times. The initial value of the effective mass $\mef^2(0^+)$ seems to be crucial for the time evolution. If $\mef^2(0^+)>0$ and $\lambda\to 0$ then $\mef(t)$ exhibits weak oscillations and the adiabaticity condition is satisfied for all times. At large times the argument of the $\cos$ increases like $2\bar{\omega}_k t$ where $\bar{\omega}_k$ is the time average of ${\omega_k}(t)$. Therefore we can apply the stationary phase method to show that the oscillations decay in time and $\mef(t)$ indeed tends to a stationary value given by
\eqq[eq4]{{m^*_{qa}}^2 = m^2 + \frac{\lambda}{2} \sum_k \left ({\frac{\omega^2_k(0) + \omega^2_{0k}}{{4 \omega_{0k} \omega_k^* \omega_k(0)}}} -\frac{1}{2 \omega_k}\right )}

If however $\mef^2(0^+)<0$, the small $k$ modes exhibit, at least at short times, exponential instead of oscillating evolution and the adiabaticity condition is no longer satisfied. The latter is also true in the marginal case $\mef^2(0^+)=0$.

Although (\ref{eq4}) is not the same as the corresponding equation of our ansatz (\ref{self-con}), they are in perfect agreement for $\lambda \to 0$ where the quasi-adiabatic approximation is correct. In the next section we will see that it is possible to construct a system of differential equations that describe the time evolution of $\mef(t)$ exactly, thus allowing us to investigate the large $\lambda$ regime.

\subsubsection{Exact time evolution equations and numerical solution}

Let us go back to the problem of a quantum harmonic oscillator with a time dependent frequency, described by the hamiltonian (\ref{timedepH}) and the equations of motions (\ref{eqm}) and start from scratch. Inspired by the adiabatic solution (\ref{eq2}), we assume a solution of the form\cite{como-87}
\eqq[sol]{\phi(t) \sim \frac{1}{\sqrt{2 \Omega(t)}} \exp{\textstyle\left (-i\int_0^t{\Omega(t') dt'}\right )}}
where $\Omega(t)$ is a suitable function that we wish to determine. Substituting into (\ref{eqm}) we find that (\ref{sol}) is the exact solution if $\Omega(t)$ satisfies the equation
\eqq[condnum]{\frac{\ddot \Omega}{2 \Omega} - \frac{3}{4}\left (\frac{\dot \Omega}{\Omega}\right )^2 + \Omega^2 = \omega^2(t)}
By comparison with the constant frequency case we can find that the appropriate initial conditions for $\Omega(t)$ are
\eq{\Omega(0)=\omega(0), \qquad \dot \Omega(0) =0}
Notice that if the derivatives of $\omega$ are much smaller than $\omega$ itself, we reproduce the quasi-adiabatic limit where $\Omega(t) = \omega(t)$ to first order. 

Taking into account the general initial conditions for $\phi(0),\pi(0)$ we have 
\eem[sol2]{\phi(t) = & \phi(0) \sqrt{\frac{\Omega (0)}{\Omega (t)}} \cos{\textstyle\left (\int_0^t{\Omega (t') dt'}\right )} + \nn \\
+ & \pi(0) \frac{1}{\sqrt{\Omega (t)\Omega (0)}} \sin{\textstyle\left (\int_0^t{\Omega (t') dt'}\right )}}
from which, using once again the initial conditions (\ref{inexpval}), we find that the equal time correlation function is 
\eem[corf2]{ \la \phi^2(t)\ra & = \frac{1}{2 \Omega(t)} \left[ 1 + \frac{(\omega(0)-\omega_{0})^2}{2 \omega(0) \omega_{0}} + \right. \nn \\
& \left. + \frac{\omega^2(0)-\omega^2_{0}}{2 \omega(0) \omega_{0}} \cos{\textstyle\left (2\int_0^t{\Omega(t') dt'}\right )} \right]}
In fact the only difference with (\ref{a1}) is that $\omega(t)$ has been substituted with $\Omega(t)$. The overall result is that instead of (\ref{eqm}) one has to solve another differential equation (\ref{condnum}). The advantage is that the former is an operator equation while the latter is an ordinary equation and it is easier to deal with real or complex-valued functions than operators, especially since we will have to solve it numerically. 

In our interacting problem, the equal time correlation function for each momentum mode $\la\phi_k^2(t)\ra$ will be given as before by (\ref{corf2}) where $\Omega_k(t)$ is also a function of $k$. Note that $\Omega_k(t)$ itself does not have to be of the form $(k^2 + M^2(t))^{1/2}$ but for large $k$ it is asymptotically equal to $\omega_k(t)$, which ensures that nothing has changed as long as the convergence of the integral in (\ref{self1a}) is concerned. 

The system of equations (\ref{corf2}), (\ref{condnum}) and (\ref{self1a}) completely determine the time evolution of the system. Although very difficult  to deal with analytically, it can be easily integrated numerically by discretizing the $(k,t)$ space and iteratively applying the following loop: 
\begin{enumerate}
	\item calculate $\Omega_k(t)$ for each $k$ from (\ref{condnum}),
	\item calculate $\la\phi_k^2(t)\ra$ for each $k$ from (\ref{corf2}),
	\item calculate $\mef^2(t)$ from the self-consistency equation (\ref{self1a}),
	\item move one step forward in time $t \to t+dt$.
\end{enumerate}

\begin{figure}[htbp]
	\centering
		\includegraphics[width=.80\columnwidth]{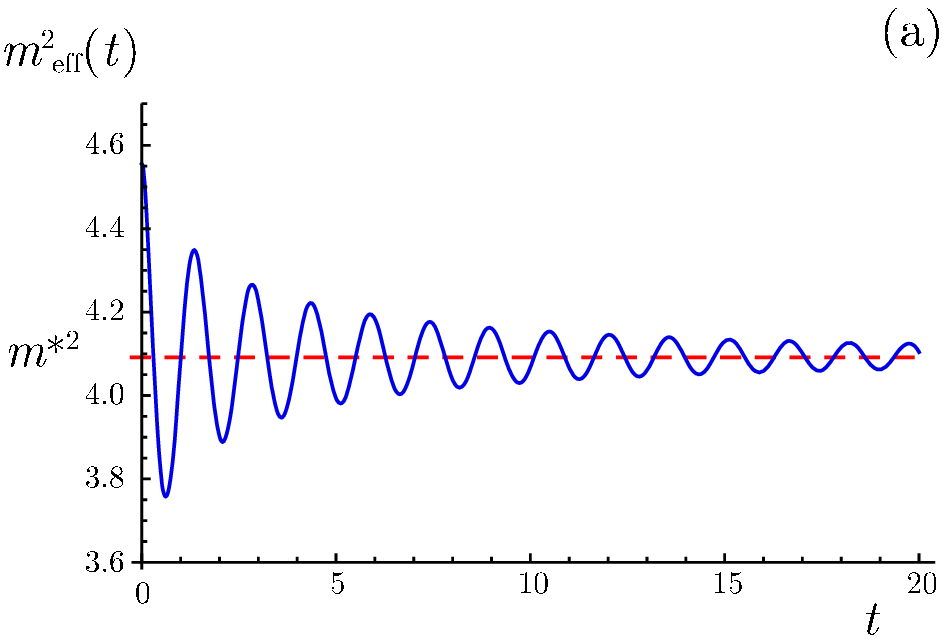}
		\includegraphics[width=.80\columnwidth]{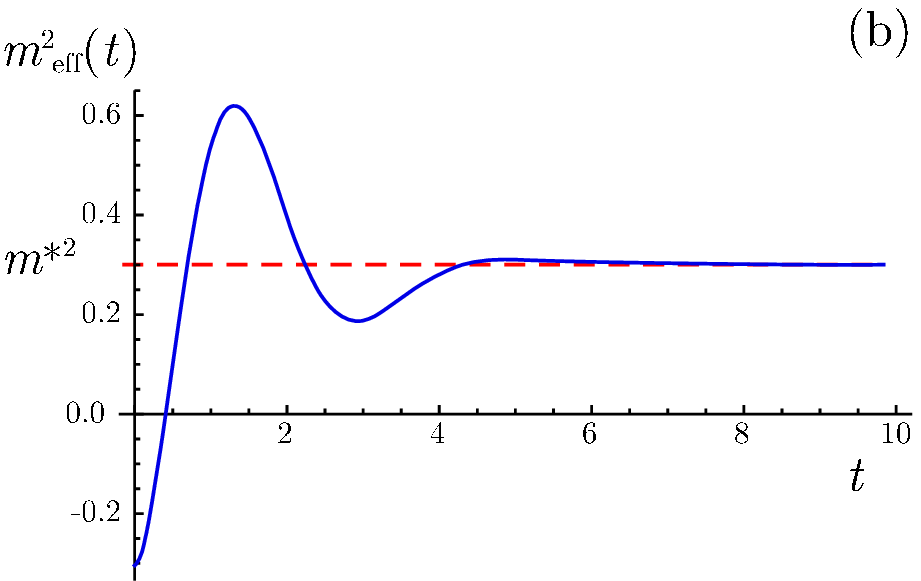}
	\caption{Typical plots of the time evolution of the effective mass as obtained numerically both in $1d$. (a) The first plot corresponds to parameter values $(m_0,m,\lambda) = (1,2,10)$ that yield a positive value for $\mef^2(0^+)$. The effective mass exhibits oscillations of decaying amplitude $\sim t^{-1/2}$ about an asymptotic value that is accurately predicted by our ansatz $m^*$. (b) The second plot corresponds to $(m_0,m,\lambda) = (1,0.5,10)$ that yield a negative value for $\mef^2(0^+)$. The initial exponential growth brings $\mef^2$ to positive values and as before $\mef$ tends to the value $m^*$ found with our ansatz. The 2$d$ and 3$d$ cases are similar. }
	\label{fig:time}
\end{figure}

Fig.~\ref{fig:time} shows typical plots of the time evolution of the effective mass. For $\mef^2(0^+)>0$ we see that the latter exhibits decaying oscillations around an asymptotic stationary value. We observe that this is the case not only for small values of $\lambda$ as we proved using the quasi-adiabatic approximation, but also for large ones. Moreover we find that even when $\mef^2(0^+)<0$ in which case the quasi-adiabatic approximation fails, $\mef^2(t)$ increases quickly and soon becomes positive to follow an oscillating evolution similar to the previously described one. The reason is that the exponential growth of the momentum modes with $k^2<-\mef^2(t)$ leads to a fast increase of $\mef^2(t)$ that brings it to positive values, ceasing the exponential growth and leaving only oscillating modes\cite{bo-93,bls-93,bdv-93}.

The asymptotic value $m^*$ as numerically estimated from the above method is systematically compared with that derived by our ansatz in the next section. It is remarkable that they are in perfect agreement for all choices of values for the parameters we studied.

\subsubsection{Comparison of the quasi-adiabatic and numerical results with our ansatz}

Let us recall our earlier ansatz for the correlation function $\tilde C(k,t)$ stating that the latter is the same, at large times, as that for a free theory with $m$ replaced by the final effective value $m^* = \mef(t\to\infty)$ which we find self-consistently, i.e.
\eem{C_{\text{ans}}(k,t) & \sim \frac{1}{2  {\omega_k^*}} \left[ 1 + \frac{( {\omega_k^*}-\omega_{0k})^2}{2  {\omega_k^*} \omega_{0k}} + \right. \nn \\
& \left. + \frac{ {\omega_k^{*2}}-\omega^2_{0k}}{2  {\omega_k^*} \omega_{0k}} \cos{\left (2  {\omega_k^* t} \right )} \right]}
On the other hand the quasi-adiabatic approximation gives
\eem{C_{\text{qa}}(k,t) & = \frac{1}{2  {\omega_k(t)}} \left[ 1 +  \frac{({\omega_k(0)}-\omega_{0k})^2}{2 {\omega_k(0)} \omega_{0k}} + \right. \nn \\
& \left. + \frac{{\omega_k^2(0)}-\omega^2_{0k}}{2 {\omega_k(0)} \omega_{0k}} \cos{\textstyle\left (2{\int_0^t{ {\omega_k(t')} dt'}}\right )} \right]}
while the exact evolution in the Hartree-Fock approximation of the problem, presented in the last section, is
\eem{C_{\text{ex}}(k,t) & = \frac{1}{2  {\Omega_k(t)}} \left[ 1 + \frac{( {\omega_k(0)}-\omega_{0k})^2}{2  {\omega_k(0)} \omega_{0k}} + \right. \nn \\
& \left. + \frac{ {\omega_k^2(0)}-\omega^2_{0k}}{2  {\omega_k(0)} \omega_{0k}} \cos{\textstyle\left (2 {\int_0^t{\Omega_k(t') dt'}}\right )} \right]}
The last two expressions differ only in that $\Omega_k(t)$ is replaced by $\omega_k(t)$ in $C_{\text{qa}}(k,t)$. An important difference between both last two expressions and $C_{\text{ans}}$ is that in the latter $\omega_k^*$ replaces $\omega_k(0)$. Furthermore although the argument of the $\cos$ in $C_{\text{qa}}$ should tend to $2 {\omega_k^* t}$ as in $C_{\text{ans}}$ for large $t$, this is not necessary for $C_{\text{ex}}$. 
Thus $C_{\text{ans}}$ is not apparently consistent with either $C_{\text{ex}}$ or $C_{\text{qa}}$, except for $\lambda\to 0$ where all of them are in agreement. 

\begin{figure}[htbp]
	\centering
		\includegraphics[width=.90\columnwidth]{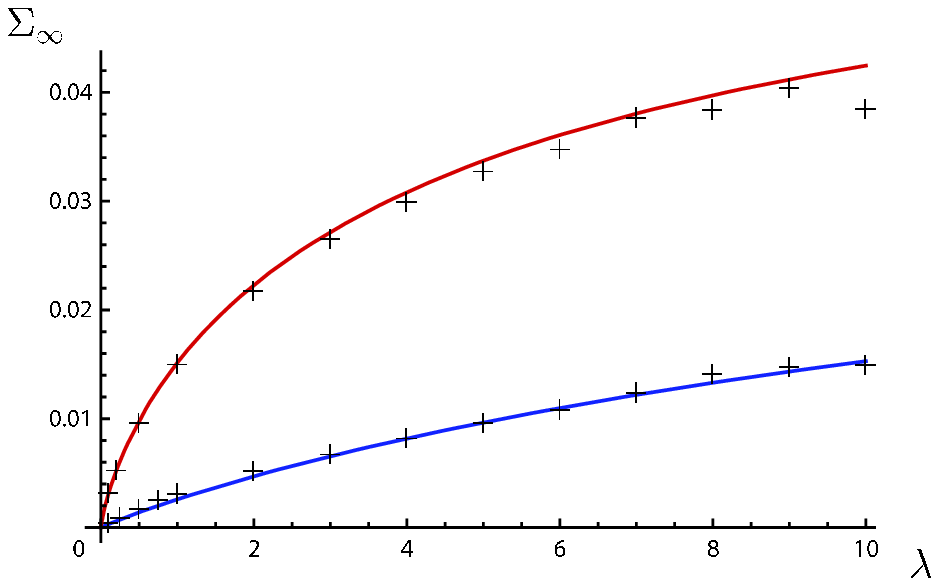}
\caption{Comparison of numerical data (crosses) with our ansatz (lines) for $m=0$. The plots are $\Sigma_\infty=\mef^2(\infty)-m^2$ as a function of $\lambda$ in units $m_0=1$. The red line corresponds to $2d$ and the blue one to $3d$ with $\Lambda=100$.}
	\label{fig:comp0}

\

\

	\includegraphics[width=.90\columnwidth]{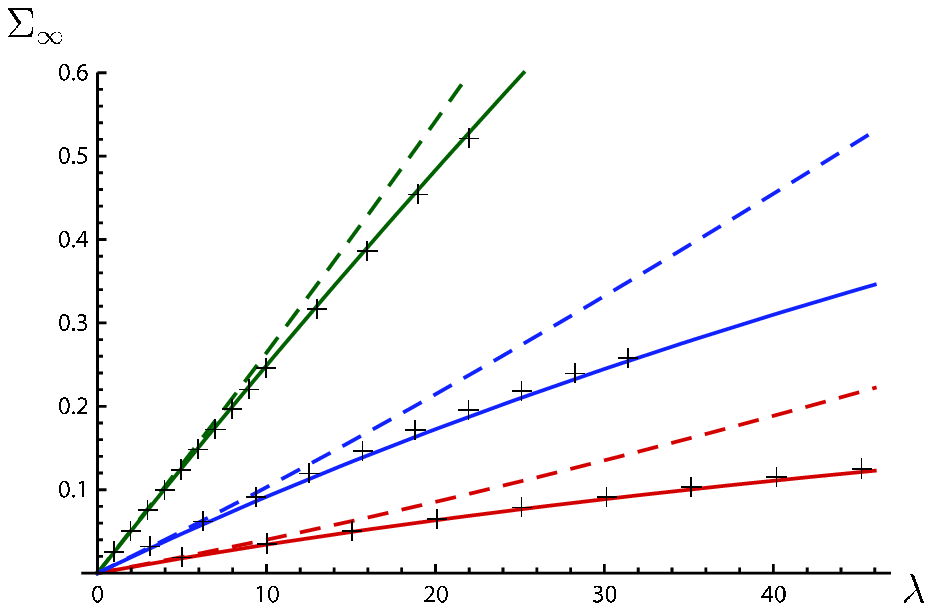}
\caption{Comparison of numerical data (crosses) with our ansatz (solid lines) and quasi-adiabatic predictions (dashed lines) for several values of $m$ (again in units $m_0=1$). The red lines correspond to $2d$ and $m=2$, the green ones to $2d$ and $m=5$ and the blue ones to $1d$ and $m=2$. It is clear that the numerics agree with our ansatz rather than the quasi-adiabatic approximation which is only good for small values of $\lambda$.}
	\label{fig:comp}
\end{figure}

Lacking an analytical argument to verify our ansatz, we rely on the numerical evaluation of the exact expression and determination of the corresponding asymptotic value of $\mef$.  Fig.~\ref{fig:comp0}, \ref{fig:comp} show plots of the shift $\Sigma_\infty=\mef^2(\infty)-m^2$ as a function of $\lambda$ for various choices of the parameter values and dimensionality, always in units $m_0=1$. The plots are based on the predictions of our ansatz, of the quasi-adiabatic approximation and estimates drawn from numerical integration of the exact equations. By comparison we observe that the numerical data agree with our ansatz very well even for large values of $\lambda$. In the contrary they do not agree with the quasi-adiabatic results, apart from first order in $\lambda$. We conclude that, although our ansatz is not manifestly consistent in form with the exact solution, it however reproduces the exact results very successfully. 

\section{Conclusions}

We studied the problem of a quantum quench in which we simultaneously change the mass and the coupling constant of an interacting system. We restrict ourselves to the time dependent Hartree-Fock approximation and make the plausible hypothesis that for large times the two-point correlation function is the same as the propagator but with a mass shift. We verify the self-consistency of our ansatz and derive the asymptotic effective mass as a function of $m, m_0$ and $\lambda$ which is shown to be correct by numerics. We point out that if $\mef(t)$ approaches its final value $\mef(\infty)$ sufficiently quickly then in the Hartree-Fock approximation the composite quench of the mass and the coupling constant is essentially nothing but a simple quench of the mass from $m_0$ directly to $\mef(\infty)$. In this case our ansatz would be justified and its generic success is probably an indication that such a fast `relaxation' process is indeed what happens. 

\begin{figure}[htbp]
	\centering
		\includegraphics[width=0.5\columnwidth]{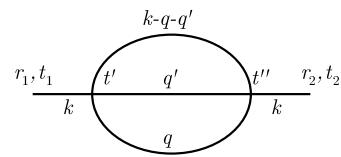}
	\caption{The `sunset' diagram.}
	\label{fig:sunset}
\end{figure}

Our findings show that effective thermalization, one of the highlights of quantum quenches in free and $1d$ conformal systems, is also possible in interacting systems such as the present model. Furthermore it is enhanced in some sense by the presence of interactions, since it occurs under more general conditions than in free systems (that is even in $2d$ massless systems). This is because of the shift in the effective mass of the system induced by the interactions. As this is their only effect in our approximation, the effective temperature is still given by the same relation as in a free model but with $m$ replaced by $m^*$, thus depending on the coupling constant. In particular, the effective temperature is still momentum dependent as in the free case, but this should not be surprising:  
as explained in the introduction and the main text, in diagrammatic perturbation theory the Hartree-Fock approximation amounts to keeping only `cactus-diagrams', i.e. Feynman diagrams that can be constructed solely by loops, and ignores the effect of collisions between quasiparticles with different momenta that can induce a mixing of the different modes. The next order correction would be to take into account the `sunset' diagram shown in Fig.~\ref{fig:sunset}.

We finally mention that, except for the stationary behaviour, also the other qualitative features of the two-point correlation function that we observed in section \ref{secpropprop} for the mass quench in the free case, are general and present also in interacting models and for quenches of the interaction strength. This comment refers not only to the horizon effect for which it is obvious, but also to the characteristic oscillations\cite{gdlp-07}, either decaying or not, and is valid at least for integrable models (or even for sufficiently small deviations from integrability) since, as the present work suggests, the first effect to the quench is only a shift of the quasiparticle masses (equivalently of the poles of the scattering matrix).

\begin{acknowledgments}
This work was supported by the EPSRC grant EP/D050952/1 and the grants INSTANS (from ESF) and 2007JHLPEZ (from MIUR). S.~Sotiriadis also acknowledges financial support from St John's College, Oxford, and the
A.~G.~Leventis Foundation.
\end{acknowledgments}

\appendix

\section{A conserved quantity}
\label{app:cons}

In the Hartree-Fock approximation the effective frequency of each momentum mode is time dependent so that the time derivative of the corresponding single mode `hamiltonian' is not zero. Indeed
\eem{\frac{d}{dt} h_k(t) & \equiv \frac{d}{dt}\left (\frac{1}{2}\dot\phi_k^2 + \frac{1}{2}\omega_k^2(t)\phi_k^2\right ) \nn \\
& = \frac{1}{2} \left \{\dot\phi_k, \ddot \phi_k + \omega^2_k(t) \phi_k \right\} + \frac{1}{2}\frac{d}{dt}(\omega^2_k(t))\phi_k^2 \nn \\
& = \frac{1}{2}\frac{d}{dt}(\omega^2_k(t))\phi_k^2 } 
where in the last step we used the equations of motion $\ddot \phi_k + \omega^2_k(t) \phi_k = 0$. 

However we can still construct a conserved quantity. From the self consistency equation (\ref{self1a}) we see that
\eq{\frac{d}{dt}(\omega^2_k(t)) = \frac{\lambda}{2} \dot C(t)}
where
\eq{C(t) \equiv \sum_{k'}{\la \phi_{k'}^2(t)\ra}}
Therefore
\eq{\frac{d}{dt} h_k(t) = \frac{1}{4}\lambda\dot C(t) \phi_k^2(t)} 
and if we take the expectation value on the initial states and sum over all momenta we conclude that
\eq{\frac{dh(t)}{dt} \equiv \frac{d}{dt}\sum_k{\la h_k(t)\ra} = \frac{1}{4}\lambda \dot C(t) C(t) = \frac{1}{8} \lambda \frac{d}{dt}(C^2(t)) }
i.e. the following quantity
\eq{h(t)-\frac{1}{8} \lambda C^2(t)}  
is conserved. As a demonstration of internal consistency, the last expression is precisely the Hartree-Fock form of the hamiltonian (\ref{Hint}) according to the substitution (\ref{subHF}).

\section{The adiabatic approximation}
\label{app:adiab}

We consider the quantum harmonic oscillator with time dependent frequency, described by the hamiltonian (\ref{timedepH}). The latter can be diagonalized in terms of the \emph{instantaneous} creation and annihilation operators $a^\dagger (t)$ and $a(t)$ defined by\cite{calhu-88}
\eq{a(t) = \sqrt{\frac{\omega(t)}{2}} \left ( \phi + i \frac{\pi}{\omega(t)} \right )}
and its hermitian conjugate. Notice that $a(t)$ in the above relation depends on time only through $\omega(t)$. The time evolution due to the dynamics of the problem is obtained from the Heisenberg equations of motion which in the case of operators that depend explicitly on time become
\eq{\frac{d a}{dt} = i[H,a] + \frac{\partial a}{\partial t} = - i \omega a + \frac{\dot{\omega}}{2 \omega} a^\dagger}
and its hermitian conjugate. The last equations form a system of linear differential equations that in matrix form looks like
\eqq[matrixeq]{
\frac{d}{dt}
\begin{pmatrix}
  a \\
  a^\dagger
\end{pmatrix} = 
A(t) 
\begin{pmatrix}
  a \\
  a^\dagger
\end{pmatrix} \; , \quad A(t) \equiv 
\begin{pmatrix}
  -i \omega & \frac{\dot{\omega}}{2 \omega} \\
  \frac{\dot{\omega}}{2 \omega} & +i \omega 
\end{pmatrix} 
}
with solution
\eqq[eq1]{
\begin{pmatrix}
  a(t) \\
  a^\dagger (t)
\end{pmatrix} = 
\mathcal{T} \exp{\textstyle\left ( \int_0^t A(t') dt' \right )}
\begin{pmatrix}
  a(0)\\
  a^\dagger(0)
\end{pmatrix}
}
where $\mathcal{T}$ denotes time ordering. 
If the frequency varies only slowly (adiabatically) with time then $\dot{\omega}/\omega^{2} \ll 1$ and $A(t)$ can be approximated by 
\eq{A(t) \approx
\begin{pmatrix}
  -i \omega & 0 \\
  0 & +i \omega 
\end{pmatrix} 
}
which is diagonal, so that the solution to (\ref{matrixeq}) is simply
\eqq[e1]{
a(t) = \exp{\textstyle\left ( - i \int_0^t{\omega(t') dt'} \right )} a(0)
}
and its hermitian conjugate. Note that the first order correction due to the off-diagonal part of $A(t)$ gives
\eem[e2]{ a(t) = & e^{ - i \int_0^t{\omega(s) ds}} a(0) + \nn \\
& + e^{- i \int_0^t{\omega(s) ds} } \int\limits_0^t{dt' \, \frac{\dot{\omega}(t')}{2 \omega(t')} e^{{ 2 i \int_0^{t'}{\omega(s) ds}}} } a^\dagger(0)}  
Keeping only the zeroth order term, we proceed to finding $\phi(t)$ from 
$\phi(t)= \left ( a(t) +a^\dagger(t) \right )/\sqrt{ 2 \omega(t)}$ 
to obtain (\ref{eq2}) in the main text.

\bibliographystyle{unsrt}
\bibliography{references}{}

\end{document}